\documentclass{aa}  

\usepackage{graphicx}
\usepackage{txfonts}
\usepackage[colorlinks=true,    % 用彩色文字而不是彩色边框
            citecolor=blue,     % \cite 链接为蓝色
            linkcolor=blue,      % \ref（公式/图表/节等）为红色
            filecolor=blue]
            {hyperref}
\setcitestyle{authoryear,open={},close={},aysep={~},citesep={, }}

\newcommand{\samethanks}[1][\value{footnote}]{\footnotemark[#1]}

\begin{document}

   \title{Radio AGN feedback sustains quiescence only in a minority of massive galaxies}

   \author{
    Huiling Liu\inst{1,2}\thanks{HLL, YL and HH contributed equally to this work.}\and 
    Yan Lu\inst{3,4}\samethanks\and 
    Hui Hong\inst{1,2,5}\samethanks\and 
    Huiyuan Wang\inst{1,2}\thanks{Corresponding author: whywang@ustc.edu.cn}\and
    Houjun Mo\inst{6}\and
    Jing Wang\inst{7}\and
    Wanli Ouyang\inst{3,4}\and
    Ziwen Zhang\inst{1,2}\and
    Enci Wang\inst{1,2}\and
    Hongxin Zhang\inst{1,2}\and
    Yangyao Chen\inst{1,2}\and
    Qinxun Li\inst{1,2,8}\and
    Hao Li\inst{1,2}\and
    Mengkui Zhou\inst{1,2}
    }

   \institute{
   Department of Astronomy, University of Science and Technology of China, Hefei, Anhui 230026, China \and
    School of Astronomy and Space Science, University of Science and Technology of China, Hefei 230026 China \and
    Center of AI for Science, Shanghai AI Laboratory, Shanghai 200233, China \and
    Department of Information Engineering, the Chinese University of Hong Kong, Hong Kong, China \and
    Department of Physics \& Astronomy, University of California, Riverside, CA, 92521, USA \and
    Department of Astronomy, University of Massachusetts, Amherst MA 01003-9305, USA \and
    Kavli Institute for Astronomy and Astrophysics, Peking University, Beijing 100871, China \and
    Department of Physics and Astronomy, The University of Utah, 115 South 1400 East, Salt Lake City, UT 84112, USA
    % \thanks{HLL, YL and HH contribute equally to this work.} \thanks{Corresponding author: whywang@ustc.edu.cn}
    }

   % \date{Received September 15, 1996; accepted March 16, 1997}

   \abstract
   {Radio active galactic nuclei (AGNs) inject a large amount of energy into their surrounding medium, which is widely believed to prevent gas cooling and maintain quiescence in massive galaxies. However, their short-lived, sporadic and anisotropic nature, together with abundant cold gas in many quiescent galaxies, challenges the universality of this mechanism. }
   {We aim to identify the population of massive quiescent galaxies that are genuinely regulated by radio AGN feedback, independently of their instantaneous radio emission, and to study their galaxy and halo properties to assess how the feedback efficacy correlates with their cold-gas reservoirs.}
   {We develop an AI classifier using galaxy optical images, trained on Q-LERGs (Quiescent Low-Excitation Radio Galaxies)   
   and non-Q-LERGs samples with a noisy-label learning strategy. This classifier distinguishes galaxies where feedback is effective (RFE), regardless of their current radio emission, from those with ineffective feedback (RFI).}
   {Galaxies categorized as RFE are all dynamically hot, whereas quiescent RFI (RFI-Q) galaxies usually have extended cold-disk components. At given stellar mass, dark matter halos hosting RFE galaxies are between four to ten times more massive than those of RFI-Q galaxies. Most RFE galaxies have little cold gas, while many RFI-Q galaxies are surrounded by substantial amounts of atomic gas. Our findings provide direct and compelling evidence that radio AGN feedback can suppress cooling over hundreds of activity cycles but that only a small fraction of massive galaxies maintain their quenched states by this process. }
   {}


   \maketitle
%
%________________________________________________________________

\section{Introduction}

Massive galaxies in the local universe are predominantly quiescent, despite residing in massive halos where the hot gas is expected to cool rapidly, potentially leading to the formation of a dense cooling gas flow that could trigger significant star formation in the absence of a heating source (\cite{Fabian1994, McDonald2018}). The lack of such a cooling gas flow in the core of observed clusters, the so-called cooling-flow problem, is a long-standing puzzle of galaxy evolution. 

% add more observation evidences
In the widely adopted “maintenance-mode feedback” scenario, radio-loud active galactic nuclei (AGNs) are considered the "missing heating source" in massive quiescent galaxies (\cite{Fabian2012, Heckman2014, Eckert2021}). 
Such an AGN can inject large amounts of energy into their surrounding medium episodically, through outbursts of mechanical energy 
carried by jets followed by longer intervals of inactive states, so that the time-averaged energy input can balance the radiative cooling 
in the core (\cite{Best2007, Sun2009, Heckman2014, Sabater2019}). 
% So called duty cycle 
Previous studies provide some evidence for this scenario (\cite{McNamara2007, Eckert2021}). 
For example, powerful radio-loud AGNs are found to be able to inflate the X-ray cavities in the central 
parts galaxy groups and clusters. The associated enthalpy in such a cavity provides a measure of 
the time-averaged jet power, and is found to be strongly correlated with the cooling loss derived from 
the observed X-ray emission.
Statistically, the total energy output of radio-loud AGNs is well-constrained and sufficient to balance 
radiative cooling in groups/clusters of galaxies (\cite{Birzan2008,Fabian2012,Eckert2021,Liu2024}).
Therefore, this feedback mechanism is thought to be the dominant process that maintains the quenching 
of star formation in massive galaxies, and has been adopted in both semi analytic models and numerical simulations
to reproduce the observed bimodal color distributions and luminosity/stellar mass functions of galaxies   
(\cite{Croton2006, Somerville2015, Wechsler2018, Weinberger2017, dave2019simba}).

Detailed studies of radio-loud AGNs have revealed their diversity. Optical narrow-line diagnostics classify these sources 
into two categories: low-excitation radio galaxies (LERGs) and high-excitation radio galaxies (HERGs) (\cite{Kauffmann2003agn, Best2005, Best2012, Sabater2019}).
HERGs, characterized by strong high-ionization lines and typically associated with radiatively efficient accretion 
(with Eddington ratio > 1\%), constitute a few per cent of the overall radio-loud AGN population in the local universe. 
In contrast, LERGs, which display weak high-ionization lines and are linked to radiatively inefficient accretion 
(with Eddington ratio < 1\%), are dominating population of radio-loud AGNs in the local universe.
Observations suggest that local HERGs and LERGs are distinctive in the properties of their host galaxies:
in general galaxies hosting HERGs tend to be less massive, with younger stellar populations and larger amounts of 
ongoing star formation (\cite{Best2012, Hardcastle2020, Hardcastle2023}). 
These results suggest that LERGs are more responsible for the maintenance-mode feedback (\cite{Hardcastle2020}).

In spite of these case studies, it remains unclear whether radio feedback can consistently and persistently maintain 
the quiescence of massive galaxies. For example, mechanisms for heat deposition and transport, such as the roles of 
sound waves, turbulence, cosmic rays and mixing, remain poorly understood (\cite{Dekel2006, Banerjee2014, Beckmann2022}).
Thus, it is still uncertain whether jets, with their short-lived, sporadic and anisotropic nature, can effectively compensate 
the gas cooling that typically has a timescale of Gyrs (\cite{McNamara2007, Birzan2008, Pinjarkar2023, Su2021}). 
%Theoretically, different possibilities have been proposed and studied \cite{Su2021, HeA2025}.
When weak radio AGNs detected by LOFAR survey are included, the fractional incidence of radio AGN activity 
seems to approach unity at the massive end ($\log(M_\star[M_\odot])>11$, with considerable uncertainty \cite{Sabater2019}). 
However, independent studies found that in geometrically flat, quiescent galaxies, there is noticeable 
absence of radio-loud AGN activity (\cite{McLure1999, Barisic2019, Zheng2022}), 
implying that the efficiency of radio AGN feedback is not uniform over the population of massive galaxies. 
For instance, the efficiency for converting baryonic gas into stars is about 50\% for massive star-forming disk 
galaxies (\cite{Posti2019, ZhangZ2022}), indicating that mechanisms to suppress star formation in these galaxies
are not very effective in quenching star formation. Moreover, many quiescent galaxies retain substantial amounts of 
cold gas (\cite{Serra2012, ZhangC2019, LiX2024}), suggesting that heating from AGNs or other sources 
cannot prevent cooling and condensation in these quiescent galaxies. 

Therefore, it is important to separate galaxies that are truly regulated by radio AGN feedback from those that are not. 
In the literature, many studies constructed radio galaxy samples by selecting galaxies with strong radio AGN activity
(\cite{Best2012, Sabater2019, Hardcastle2025}) and used them to study the duty cycle and the efficiency of radio feedback. 
However, non-detection in radio activity does not mean a lack of regulation because previous AGN outbursts may already have heated and/or expelled 
the surrounding gas so that the production of cold gas is suppressed. If maintenance-mode heating can indeed maintain 
quiescence through repeated activity cycles, it may leave long-lived and learnable statistical imprints on the stellar 
structure and population captured in optical images, to distinguish them from systems with ineffective radio feedback. 
Additionally, because the duty cycle of radio-loud AGNs is much shorter than the evolution timescale of the stellar 
population and galactic structure (\cite{Birzan2008, Turner2015, Pinjarkar2023, Benjamin2025}), these imprints 
are expected to persist even when supermassive black holes are currently inactive. 

%These reveal a conceptual gap: how can we identify galaxies that are truly regulated by radio AGN feedback? A practical complication is that most observational diagnostics track only the current state of AGN (\cite{Best2012, Sabater2019, Hardcastle2025}). Even with an inactive SMBH at present, prior AGN outbursts may have supplied maintenance-mode feedback, so current non-detection in the radio band does not imply a lack of regulation.  
% Moreover, this history is not always recoverable from X-ray cavities alone, given their time evolution and detectability limits. 
%If maintenance-mode heating indeed maintains quiescence across repeated activity cycles, it should leave long-lived, subtle but learnable statistical imprints in the stellar structure, environment, merger signatures, and dust morphology captured in optical images, distinct from systems with ineffective radio feedback. Because the duty cycle of radio-loud AGN is much shorter than a galaxy’s dynamical time (\cite{Birzan2008, Turner2015, Pinjarkar2023, Benjamin2025}), these imprints should persist even when the AGN is currently inactive.

This motivates the search for a classification to distinguish between galaxies where maintenance-mode feedback has been 
crucial in maintaining quiescence and those where radio AGNs do not impact the host galaxies significantly. 
We refer to the former as radio-feedback-effective galaxies (RFE) and the latter as radio-feedback-ineffective galaxies (RFI).
The RFE galaxies, which are quiescent because of maintenance-mode feedback, comprise two subgroups: 
galaxies with ongoing powerful AGN activity (termed RFE-on galaxies) and those with weak or undetectable current 
AGN activity (RFE-off galaxies). It is possible that an RFI galaxy hosts a radio AGN but it fails to maintain its quiescence. 
Thus the RFI population also includes two subcategories, namely, RFI star-forming galaxies (RFI-SF) and RFI quiescent galaxies (RFI-Q).  
In general, it is difficult to use conventional method to distinguish between RFE-off and RFI-Q galaxies because they are 
both quiescent and undetected as radio AGNs.

\begin{table*}
\caption{Terminologies and samples used in this paper}      
\label{tab_sample}      
\centering          
\begin{tabular}{l l l} 
\hline\hline       
Name & Description & Number \\
\hline
& Classification using radio activity and star formation&\\\hline
Q-LERG &  quiescent Low-Excitation Radio Galaxies  & 3,095\\
non-Q-LERG & galaxies not classified as Q-LERG & 397,214 \\
\hline\hline
& Feedback efficacy classification using AI&\\\hline
RFE &   Radio-feedback-effective galaxies  &  \\
RFE-on & Q-LERGs, RFE galaxies with powerful radio activity & 3,095\\
RFE-off & RFE galaxies with undetectable AGN activity& \\
% Radio-feedback effective galaxies with turned-on radio AGN, 
% quchened LERGs\cite{Best2012}, excluded if $^0.1(g-r)<0.7$
pRFE-off & High purity RFE-off sample & 8,693 \\ 
cRFE-off & High completeness RFE-off sample & 38,784 \\
pRFE & RFE-on and pRFE-off & 11,788\\ 
cRFE & RFE-on and cRFE-off & 41,879\\
%\hline
RFI & Radio-feedback-ineffective galaxies& 358,430\\
RFI-Q & Quiescent RFI galaxies & 121,442\\
RFI-SF & Star-forming RFI galaxies & 236,988\\ 
\hline
\hline       
\end{tabular}
\tablefoot{All galaxies are central galaxies of galaxy groups.
Q-LERGs and non-Q-LERGs are classified by using the radio activity and SFR. The non-Q-LERG sample comprises cRFE-off and RFI samples. 
We utilized the AI to classify galaxies into RFE (RFE-on$+$RFE-off) and RFI(RFI-Q$+$RFI-SF) samples. 
The RFE-on sample is exactly the same as the Q-LERG sample.
Because of the difficulties discussed in the text, we actually obtained two 
RFE-off samples, pRFE-off(higher purity) and cRFE-off(higher completeness), and therefore two RFE samples, pRFE and cRFE. Note that the pRFE-off (pRFE) sample is the subset of the cRFE-off (cRFE) sample.}
\end{table*}

In this work, we develop an AI classifier that operates on optical images of to identify 
galaxies for which radio feedback is effective in sustaining quenching, independent of their radio emission at the present. 
We train the classifier using traditional labels, with quiescent low-excitation radio galaxies (Q-LERGs, \cite{Best2012}) as positives 
and non-Q-LERGs as negatives. Crucially, we introduce a noisy-label learning algorithm (\cite{wang2019symmetric}) because 
(i) Q-LERGs and other feedback-effective systems can be morphologically similar in the optical image, 
and
(ii) many of these other systems are radio-faint/inactive and thus labeled “non-Q-LERG” by radio-based schemes, 
which introduces noise into the negatives. 
The paper is organized as follows. 
In Section~\ref{sec:sample}, we describe the multi-band data sets used in this work.
Section~\ref{sec:classification} details the AI classification of RFE galaxies and its performance. 
Section~\ref{sec:galaxy fractions} investigates the RFE galaxy fraction and its duty cycle.
Section~\ref{sec:gal props} analyzes the difference between RFE and RFI galaxies in both halo and galactic properties.
Section~\ref{sec:gas} examines and compares the HI masses of different populations of galaxies.
Section~\ref{sec:conclusion} summarizes our results and discusses their implications.
We adopt a standard flat cosmology with $H_0 = 70~\mathrm{km\,s^{-1}\,Mpc^{-1}}$, $\Omega_m = 0.3$, and $\Omega_\Lambda = 0.7$ throughout the paper.

%and the radio band spectral index $\alpha=-0.7$, which is defined in the sense $S \propto \nu^{\alpha}$. {\color{red}move this sentence to the proper section.}

%__________________________________________________________________

\section{Sample and data}
\label{sec:sample}

\begin{figure*}
  \centering
  \includegraphics[width=0.85\textwidth]{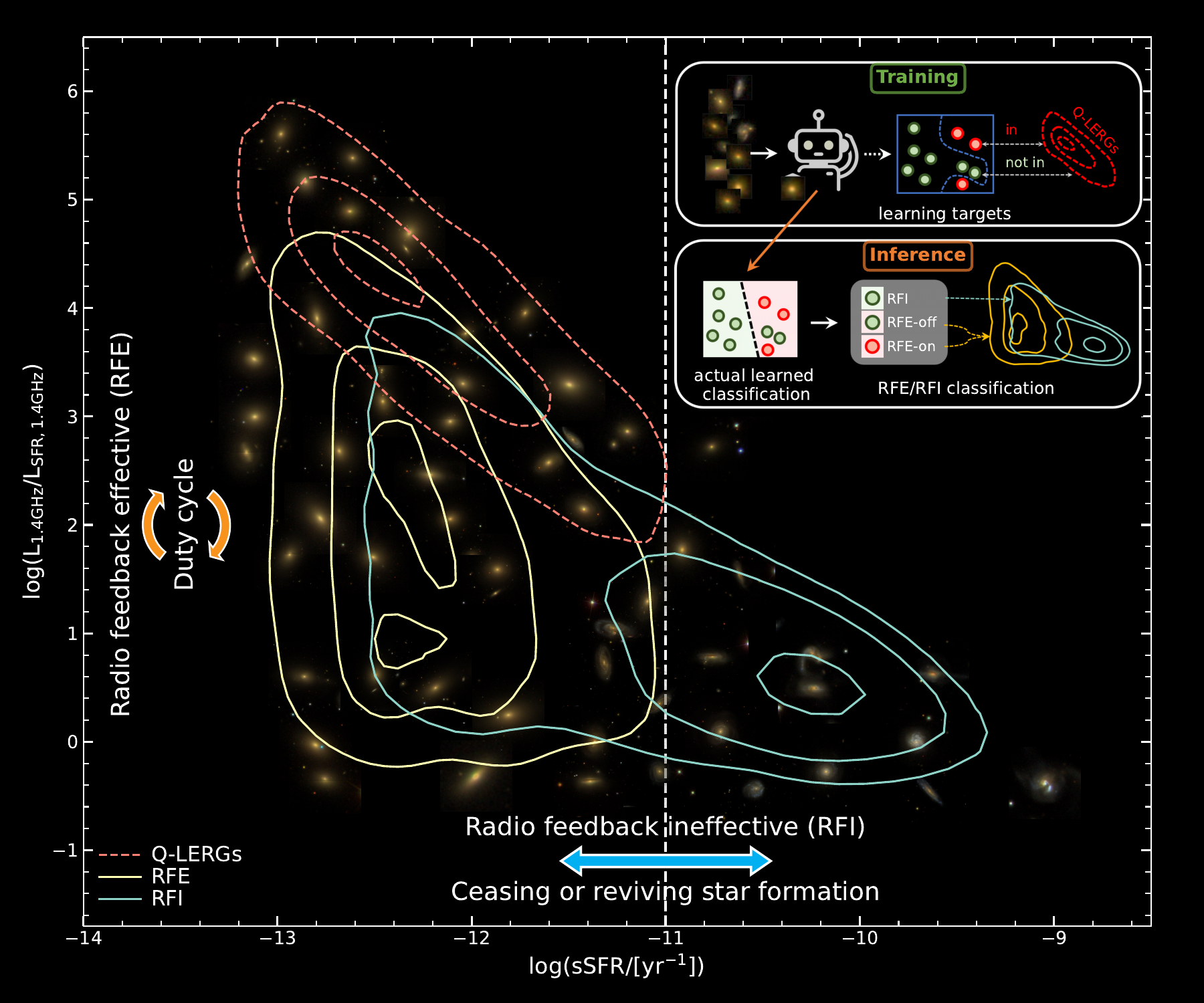}
  \caption{
  \textbf{AGN-SF diagram and AI pipeline.}
    The observed radio luminosity ($L_{\rm 1.4GHz}$) scaled by the predicted radio luminosity ($L_{\rm SFR,1.4GHz}$) from the star-formation rate(SFR) versus the specific SFR (sSFR). 
    The red dashed contours exhibit Q-LERGs, a representative sample of RFE-on galaxies. The cyan and yellow contours show the RFI and RFE (cRFE) populations, which are the outcome of the AI analysis.
    The contour lines enclose 10\%, 50\% and 90\% of the galaxies with $\log M_\star/M_\odot>10.5$.
    The vertical dashed line indicates the demarcation between quiescent and star-forming galaxies. 
    RFE galaxies may move vertically as they enter the duty cycle of AGN activity on a short timescale, while RFI galaxies can move horizontally as their SFRs evolve (usually with a longer timescale). 
    We show the images of randomly selected galaxies in the background. Most star-forming galaxies are blue and exhibit spiral arms; 
    most galaxies with large $L_{\rm 1.4GHz}/L_{\rm SFR,1.4GHz}$ are red and appear to be elliptical galaxies; 
    both elliptical and disk-like galaxies exist in the lower-left section of the diagram. 
    The RFE classifier, which is trained by Q-LERG labels only, can classify RFE and RFI based on SDSS galaxy images. 
    This is achieved by combining both training the model to classify Q-LERG 
    (RFE-on) versus non-Q-LERG (RFE-off + RFI) and deliberately encouraging the model to misclassify RFE-offs 
    into the Q-LERG category through a noise-label learning scheme.    
    Since RFE-off and Q-LERG galaxies share similar optical appearance in the image domain, we 
    successfully generated an RFE classifier that distinguishes RFE (RFE-on + RFE-off) from RFI. See Table~\ref{tab_sample} for the introduction of the terminologies and samples used in this paper.
  }
  \label{fig:AGNSF}
\end{figure*}

\subsection{Galaxy sample}

Our galaxy sample is taken from the New York University Value Added Galaxy Catalog sample (NYU-VAGC, \cite{Blanton-05a}) of the Sloan Digital Sky Survey DR7 (SDSS DR7, \cite{Abazajian-09}). Galaxies with r-band Petrosian magnitudes $r\le$ 17.72, with redshift completeness $>$ 0.7, and with redshift range of $0.01 \le z \le 0.2$, are selected.
We only focus on central galaxies, defined as the most massive galaxies within dark matter halos, identified according to the halo-based group-finding algorithm (\cite{Yang2005, Yang2007}). 
The NYU-VAGC catalog also provides spectroscopic velocity dispersions $\sigma_\star$, rest-frame colors $^{0.1}\rm{(g-r)}$, which are k-corrected to $z = 0.1$, and measurements of the sizes of each galaxy, $R_{50}$ and $R_{90}$,
which are the radii enclosing 50 and 90 percent of the Petrosian $r$-band flux of the galaxy, respectively. 
The central velocity dispersion $\sigma_{\rm c}$ within $R_{50}/8$ is estimated from $\sigma_\star$ using $\sigma_{\rm c}=\sigma_{\star}(R_{50}/8R_{\rm ap})^{-0.066}$, where $R_{\rm ap}=1.5''$ is the size of SDSS spectroscopic fiber (\cite{Bezanson2011}). 
The mass of the central supermassive black holes is estimated(\cite{Saglia2016}) as,
\begin{equation}
    \log M_{\rm BH}[M_\odot]= 5.246 \log \sigma_{\rm c}[{\rm km/s}] - 3.77\\.
\end{equation}
The concentration of a galaxy is defined as $c=R_{90}/R_{50}$, usually used to indicate the morphology of the galaxy (\cite{Shimasaku2001}).

We cross-match our sample with the MPA-JHU DR7 catalog to obtain the measurements of the stellar mass, $M_\star$, which is derived by fitting the SDSS $ugriz$ photometry (\cite{Kauffmann2003b}).
The star-formation rate (SFR) of our galaxies is taken from the GSWLC-2 version (\cite{Salim2016}), obtained by fitting the UV$+$optical$+$mid-IR SEDs. 
We adopted $\log(\rm{sSFR}[\rm{yr}^{-1}]) = -11$ to separate quiescent galaxies from star-forming galaxies, where sSFR=SFR/$M_\star$.  
The axis ratios, which describe the projected ellipticity of galaxies, are obtained by cross-matching our sample with the \verb|PhotoObjAll| table of SDSS DR12 (\cite{SDSSDR12}). 
%The axis ratio a/b of a galaxy, describing its projected ellipticity, is characterized by either \verb|deVAB_r| or \verb|expAB\_r|, which are obtained from fitting a de Vaucouleurs or an exponential profile, respectively. Based on the \verb|fracDeV_r| parameter, which indicates the relative contribution of a de Vaucouleurs component in the composite model, we adopted axis ratio \verb|deVAB_r| for galaxies with \verb|fracDeV_r|  $>$  0.5 and \verb|expAB_r| otherwise. 

We downloaded the SDSS optical images for these galaxies from the SDSS website.  Each galaxy image is positioned at its center, with a size equal to 6 times $R_{50}$ (512 pixels on each side).  
These images are constructed from SDSS data in the $g$, $r$ and $i$ bands to form 3-color jpg images using the published pipeline (\cite{Lupton2004PASP}).  
In the following, we use galaxy images for AI training.
The image size is chosen as stated above to ensure that both the central and peripheral regions of galaxies are included in the AI learning process, while minimizing the influence of other galaxies or objects. 
%Because the image size is proportional to the galaxy size, the images do not carry much information about the redshift, stellar mass, and physical size of individual galaxies.  However, the concentration of light distribution can be derived directly from the images.
Our final central galaxy sample consists of 400,309 galaxies, of which 259,148 have $M_\star\ge 10^{10.5}M_\odot$.

\subsection{Radio sources and AGN-SF diagram}

We cross-matched our galaxy sample with the local radio source sample from \cite{Best2012}, which has been widely used to study the properties, distributions, and evolution of radio galaxies (see e.g. \cite{Heckman2014}). This radio source sample was assembled from the SDSS DR7 and the National Radio Astronomy Observatory's Very Large Array Sky Survey (NVSS, \cite{Condon1998}), as well as the Faint Images of the Radio Sky at Twenty Centimeters survey (FIRST, \cite{Becker1995}). A total of 5,506 central galaxies are detected at 1.4GHz. 
According to \cite{Best2012} , 3,797 central galaxies are classified as radio AGNs. There are two fundamentally different radio AGN classes, `Low-Excitation Radio Galaxies' (LERGs) and `High-Excitation Radio Galaxies' (HERGs). 
About 3\% of radio galaxies are classified as HERGs, and 10\% are unclassified.
HERGs have Eddington ratios between one and ten percent and are usually associated with younger stellar populations, indicating that they are unlikely to be responsible for maintaining long-term quiescence in their host galaxies. 
In contrast, LERGs exhibit lower Eddington ratios (mostly below 1\%) and tend to reside in more massive galaxies with older stellar population (\cite{Best2012}).
Since our purpose is to investigate the role of the maintenance-mode AGN feedback, which is believed to be capable of maintaining the quiescence of its host galaxy, we consider only quiescent LERG labeled Q-LERG. Almost all of these galaxies are red, and 6 galaxies with blue color ($^{0.1}\rm(g-r)<0.7$) are excluded. There are 3,095 Q-LERGs, and the remaining 397,214 galaxies in our sample are labeled non-Q-LERGs. 
About 93\% of the Q-LERGs have radio luminosity at 1.4 GHz ($L_{\rm 1.4GHz}$) 
higher than $10^{23}\rm W~Hz^{-1}$. Therefore, the Q-LERG sample is dominated by radio galaxies with powerful radio activity. 
Should radio feedback truly suppresses gas cooling, it must show efficacy in the Q-LERG sample. 
We thus use the Q-LERG sample as a representative sample for RFE-on galaxies. The terminologies and samples used in this paper are listed in Table~\ref{tab_sample}.

%\subsection{AGN-SF diagram}

%\begin{figure*}
%\includegraphics[width=0.85\linewidth,keepaspectratio]{Figs/figex_LNVSS_SFR.pdf}
%\centering
%\caption{{\bf The 1.4 GHz Radio luminosity-SFR relation.} 
%        \textbf{A}: The $L_{\rm{1.4GHz}}$ versus SFR relation for galaxies with $\log M_\star/M_\odot>10.5$. The radio data are taken from \cite{Best2012}. The background shows the 2D hexagon histogram of detected galaxies in \cite{Best2012} with number indicated by the color bar. The black line shows the $L_{\rm{SFR,1.4GHz}}$-SFR relation for star-forming galaxies (\cite{Davies2017}). The red contours show the distribution for Q-LERGs, orange for all radio-detected galaxies, blue for  radio-undetected galaxies with $L_{\rm 1.4GHz}$ being the upper-limit, and green for radio-undetected galaxies with $L_{\rm 1.4GHz}$ being a randomly assigned value between the upper-limit and the prediction based on the SFR taking into account scatter. The contour lines enclose 10\%, 50\%, 90\% of the corresponding samples. \textbf{B}: the same as in \textbf{A}, but shown in the same parameter space as Fig.~\ref{fig:AGNSF}: $L_{\rm{1.4GHz}}/L_{\rm{SFR,1.4GHz}}$ versus sSFR, where $L_{\rm{SFR,1.4GHz}}$ is the expected radio luminosity from star formation\cite{Davies2017}. } \label{fig:LNVSS_SFR}
%\end{figure*}

The AGN-SFR diagram in Fig.~\ref{fig:AGNSF} shows $L_{\rm 1.4GHz}/L_{\rm SFR, 1.4GHz}$ versus the specific star formation rate(sSFR). 
Here, $L_{\rm SFR, 1.4GHz}$ is the expected 1.4GHz luminosity powered by star formation (\cite{Davies2017}):
\begin{equation}
    \log(L_{\rm{SFR, 1.4GHz}}[\rm{W/Hz}]) = 1.52 \log(\rm{SFR[M_\odot/yr]}) + 21.24
\end{equation}
Both galaxies detected and undetected in radio emission are shown in this figure. We assign a mock $L_{\rm 1.4GHz}$ to radio-undetected galaxies as follows. We first calculate the upper limit ($L_{\rm lim}$) of $L_{\rm 1.4GHz}$ for these galaxies. $L_{\rm lim}$ is derived from the flux limit (\cite{Best2012}) of 5 mJy and the redshift of the galaxy
\begin{equation}
    L_{\rm{lim}} =4\pi D_{\rm{L}}^2(z) F_{\rm{lim}} (1+z)^{-\alpha-1}
\end{equation}
where $D_{\rm{L}}$ is the luminosity distance and $\alpha=-0.7$ is the typical spectral index (\cite{Condon2002}). For each of these galaxies, we then generate a radio luminosity of $L_{\rm{SFR, pred}}$ based on its SFR using the $L_{\rm{SFR,1.4GHz}}-\rm{SFR}$ relation with a dispersion of 0.23 dex (\cite{Davies2017}) .
We then randomly pick a mock luminosity from a uniform distribution between $L_{\rm lim}$ and $L_{\rm{SFR, pred}}$ and assign it to the galaxy. 
Only 0.5\% of the radio-undetected galaxies have $L_{\rm{SFR, pred}}>L_{\rm lim}$. 
For those galaxies we use $L_{\rm{lim}}$ as their mock radio luminosity. 

As one can see, nearly all galaxies with prominent radio AGN activity (in the upper-left section) are quiescent. This is consistent with the widely accepted picture that the feedback from radio AGNs can suppress the cooling and condensation of gas needed for star formation and maintain the quiescence of massive galaxies (\cite{Croton2006, dave2019simba, Su2021}). 
Nevertheless, only a small fraction of quiescent galaxies show strong radio emissions (\cite{Best2012}). 
Most quiescent galaxies, which are either very faint or undetected in radio, are clustered in the lower-left section of the diagram. 
Numerous studies attributed this to the radio AGN duty cycle (\cite{Best2007, Heckman2014, Sabater2019}), in which even quiescent galaxies lacking radio AGNs today have previously been influenced by radio feedback from past AGN events. 

However, RF-Q galaxies occupy the same region in the AGN-SFR diagram as RFE-off galaxies.
Section \ref{sec:classification} will introduce an AI method for separating these galaxies. In order to evaluate the performance of our method, we introduce a deeper radio survey, the second date release from LOFAR Two-Metre Sky Survey (LOTSS DR2 \cite{Hardcastle2023}), observed at a central frequency of 144 MHz.  
LOTSS DR2 consists of 841 LOFAR pointings, covers about 5700 $\rm deg^2$ in the northern sky. We select a regular region with right ascension between 150$^{\circ}$ and 230$^{\circ}$, and declination between 30$^{\circ}$ and 65$^{\circ}$, located within the SDSS regions. A total of 33,932 of our galaxies are detected and 56,234 are not.

\subsection{MaNGA}

The SDSS Mapping Nearby Galaxies at Apache Point Observatory data release 17  (MaNGA DR17, \cite{Bundy2015})  provides spatially resolved spectra of about
10,000 galaxies in the SDSS region (\cite{Abdurrouf2022}). 
%The main NaNGA sample was selected to have a flat stellar mass distribution. 
We adopted the data product of Pip3D (\cite{Sanchez2016RMxAA2}), which provides the measurement of stellar velocity dispersion ($\sigma_*$) and the 4000\AA\ break (D4000) on each spaxel. Following our previous study (\cite{HongH2023}), we calculated the inner stellar velocity dispersion ($\sigma_{\rm in}$), outer stellar velocity dispersion ($\sigma_{\rm out}$), inner D4000 ($D4000_{\rm in}$) and outer D4000 ($D4000_{\rm out}$). The two inner parameters are the median values of the spaxels within the inner 0.2$R_{50}$ elliptical annuli, and the two outer parameters are the median within the 1.4-1.5$R_{50}$ elliptical annuli. The position angle and ellipticity that determine the elliptical annuli are taken from the NSA catalog (\cite{Blanton2011AJ}). To obtain reliable measurements of these quantities, we exclude spaxels with a signal-to-noise ratio of $\sigma_*$ less than 3. %We also exclude galaxies with QCFLAG equal to 1 or 2, or galaxies without NSA redshift or $R_{50}$ measurement.
After cross-matching with our central galaxy sample, we have a total of 5,462 galaxies with spatially resolved data. Among them, 2,789 galaxies have $\log M_\star/M_\odot>10.5$.

\subsection{xGASS}
The HI data are collected from the xGASS representative sample, provided by the The extended GALEX Arecibo SDSS Survey (xGASS, \cite{Catinella2018}). The xGASS representative sample includes randomly-selected galaxies from the parent sample of objects in the SDSS DR7 spectroscopic catalog, located within the footprint of the Arecibo Legacy Fast ALFA survey (ALFALFA, \cite{Haynes2018}). 
They are uniformly selected across the stellar mass range of $9 < \log(M_{\star}/M_\odot) < 11.5$. These galaxies were observed at 21cm by Arecibo until HI was detected or until an upper limit of a few percent in the gas fraction  ($M_{\rm HI}/M_{\star}$) was reached. 
Since most LERGs have $\log M_{\star}/M_\odot>10.5$, our analysis is restricted to galaxies above this mass limit. 
By cross-matching our central galaxy sample and the xGASS representative sample, we selected 323 galaxies that have xGASS HI data, with a redshift range of  $0.01\leq z\leq 0.05$ and $\log(M_{\star}/M_\odot) \geq 10.5$. Among these galaxies, 130 have only an upper limit for the gas fraction.

\subsection{DECaLS}

The shear catalog used to measure weak lensing signals is constructed from the Dark Energy Camera Legacy Survey Data Release 8 imaging data (DECaLS DR8 \cite{Dey2019AJ,Zou_2019}). The shape of each galaxy is measured using the FOURIER\_QUAD pipeline, which has been demonstrated to provide accurate shear measurements even for extremely faint galaxy images with signal-to-noise ratios (SNR) below 10 (\cite{Zhang2015JCAP,Zhang_2019,Wanghr2021,ZhangJ2022}). The shear catalog spans over ten thousand square degrees in the $g$, $r$, and $z$ bands, containing approximately 99, 111, and 116 million distinct galaxies in each band, respectively. We note that the images of the same galaxy on different exposures are counted as different images in the FOURIER\_QUAD method. In this work, we only use the $r$ and $z$ band data. The DECaLS region overlapped with SDSS DR7 is 4744 square degrees, corresponding to approximately $68\%$ of the SDSS survey area.

%__________________________________________________________________

\section{Identification of radio feedback effective and ineffective galaxies}
\label{sec:classification}

Galaxies can be categorized 
as Q-LERG and non-Q-LERG(Table~\ref{tab_sample}), according to their radio emission and star formation rate. A similar classification was often adopted in previous studies(\cite{Best2012, Sabater2019, Hardcastle2025, Brinchmann2004, Renzini2015}). 
This type of classification does not fully capture the nature of certain galaxies. For example, some galaxies that are influenced by radio AGNs and currently show no radio emission because of the sporadic nature of radio activity are classified as non-Q-LERG. 
These galaxies, referred to as RFE-off galaxies, along with the Q-LERGs, 
also known as RFE-on galaxies, should both be considered as RFE galaxies. Studies based on the synchrotron emission of radio jets and X-ray cavities in the intracluster medium found that a single radio-AGN activity lasts for about $10^6$ to $10^8$ years (\cite{Birzan2008, Pinjarkar2023, Benjamin2025}), a duration significantly shorter than the timescale for stellar evolution and change in galaxy morphology.  
Thus, RFE-off and RFE-on galaxies are expected to have similar optical properties. Such similarity is the key we use to separate the two populations, RFE and RFI. 
Conversely, the remaining non-Q-LERGs, which are fundamentally different from RFE galaxies, should be placed in the RFI category. 

To address the aforementioned question, we introduced an advanced noise-label learning technique, commonly used in the AI community, to analyze optical images of Q-LERGs from \cite{Best2012}, a representative sample of RFE-on galaxies, and used the training results to identify RFE-off galaxies from the complementary population (non-Q-LERGs) again based on their optical images. 
Its principle is the data semantic appearance consistency. 
For instance, in a noisy cat-dog binary classification dataset, mislabeled dog images, whose real labels are cat, often share similar semantic appearances with most cat images, which provides cues for the AI model to recognize and correct mislabeling during training. 
Here, the SDSS dataset shares similar traits. 
``Q-LERG'' and ``non-Q-LERG'' annotations can be seen as a noisy version of ``RFE'' and ``RFI'' labels, respectively.
RFE-off galaxies can be seen as mislabeled data whose labels are ``RFI'' but should be ``RFE''. 
Also, these mislabeled RFE-off galaxies are essentially RFE galaxies, so they have optical features similar to ``RFE'' galaxies shown in SDSS images, thus providing cues for the AI model to recognize them.
Motivated by this, we utilize the symmetric cross-entropy loss function (\cite{wang2019symmetric}), 
a simple yet effective general noise-label learning method, to train our RFE galaxy classifier. 
This method allows the model trained on a noise dataset containing random label errors 
to converge statistically to the result expected from training on a clean dataset(\cite{wang2019symmetric}). 
Using such a method, our classifier learned on the SDSS dataset with the Q-LERG and non-Q-LERG 
labels becomes a reliable RFE galaxy classifier, which allows us to recognize RFE-off samples. 

The entire pipeline of our AI method is shown in Fig.~\ref{fig:AGNSF}. Each SDSS image is sent to an RFE Galaxy Classifier model to predict the probability for it to be an RFE galaxy. A Noise-Label Learning 
part supervises the model with Q-LERG and non-Q-LERG labels and guides it to converge to a reliable and 
robust RFE galaxy classifier. After the training, we treat the non-Q-LERG galaxies classified as 
RFE by the model as RFE-off galaxies. In the following, we introduce the details of each part of our model.

% Still, our analysis showed that the model performs well in uncovering feedback properties of galaxies despite these difficulties, clearly demonstrating the potential of the method for solving similar problems. 

\subsection{RFE Galaxy Classifier}

The RFE Galaxy Classifier is formulated as a function $\hat{y}=f(x)$ 
where $f$ is a model that uses $x$, a given SDSS image, to predict the probability $\hat{y}$
for the given image to be an RFE galaxy. 

We set $f$ by the Convolutional Neural Network (CNN) as follows:
\begin{equation}
    \hat{y}=f(x) \rightarrow \hat{y}=\text{Cls}(\text{CNN}(x)),
\end{equation}
where $\text{CNN}(\cdot)$, implemented as ResNet50 (\cite{he2016deep}),  takes 
image $x$ and returns a feature vector $h\in\mathbb{R}^{C}$ 
($C$ is the dimension of the feature vector) obtained by using multiple stacked convolutional layers, 
max-pooling layers, skip-connection operators and activation functions. 
%{\color{red}: the above is unclear: is $x$ or $x_i$ the image? what does `by multiple' mean? 
%where does is the right parenthesis of ($C$?  what does `feedback a feature vector'? Does it mean 
%`returns a feature vector'?} 
%ResNet50~\cite{?} which achieves great success in general image processing and analysis~\cite{?}.  
%Specifically, for the image $x\in\mathbb{R}^{W\times H\times C}$ ($W$, $H$ and $C$ are width, height and channel of $x$.), ResNet can treat it into a feature vector $h\in\mathbb{R}^{C'}$ ($C'$ is the dimension of the feature vector~\cite{?}) by multiple stacked convolutional layers, max-pooling layers, skip-connection operators and activation functions. 
%
The feature vector $h$ is a low-dimensional representation of the image $x$ that is suitable for the following classifier $\text{Cls}(\cdot)$ to predict $\hat{y}\in[0,1]$. 
$\text{Cls}(\cdot)$ is implemented by a fully-connected neural network layer and a Sigmoid activation function. 
Obviously, the parameters of Cls and CNN determine the function of the entire AI model. 
The learning of the entire model is therefore to adjust the parameters of Cls and CNN, and we utilize a 
noise-label learning technique to achieve this. 

% {\bf \textit{Noise-Label Learning.} }
The learning of the RFE galaxy classifier can be defined as an optimization process and written as follows:
\begin{equation}
\label{eq:ai_erm}
    \theta^* = \arg {\rm min}_\theta\  \mathbb{E}_{(x_i,y_i)\in\mathcal{T}} \mathcal{L}(y_i,\hat{y_i}),
\end{equation}
where $x_i$ is the $i$-th sample in the dataset $\mathcal{T}$ while $y_i$ and $\hat{y}_i$ are its label and prediction, respectively. $\theta$ is the parameter set of $f$, specifically, the parameters of CNN and Cls. 
$\mathcal{L}$ is the loss function that measures the gap between the predicted probability 
$\hat{y}$ and the corresponding ground-truth label $y$. The final goal is to find an optimized parameter 
set $\theta^*$ that can minimize the expected value of the loss function. Here, we use the Symmetric Cross-Entropy loss function (\cite{wang2019symmetric}) 
for $\mathcal{L}$. The detailed formulation is as follows:
\begin{equation}
    \mathcal{L}(y,\hat{y})=\rm{CE}+\rm{RCE}
\end{equation}
Here CE represents the Cross-Entropy term. It is used to constrain the model to adapt $\hat{y}$ as close to $y$ as possible, guiding the model to learn correlations between the input image $x$ and the label $y$:
\begin{equation}
    \rm{CE}=-y\cdot \log(\hat{y})-(1-y)\cdot \log(1-\hat{y})
\end{equation}
While RCE represents the Reverse Cross-Entropy term, a noise-robust learning term, which can make the entire model converge to the same optimized results as it was trained on a clean dataset with the cross-entropy loss function  
\begin{equation}
    \rm{RCE}=-\hat{y}\cdot \log(y)-(1-\hat{y})\cdot \log(1-y)
\end{equation}
The RCE term thus allows the model to ignore the impact of noise labels in the training data (\cite{wang2019symmetric}) and give a reliable RFE galaxy classifier.

% {\bf \textit{PK-Sampling for balanced training.}} 
The Q-LERGs and non-Q-LERGs exhibit a severe class imbalance, with 3,095 positive samples (Q-LERGs) and 397,214 negative samples (non-Q-LERGs). Direct training on this imbalanced dataset would result in a trivial classifier biased toward the majority class (i.e., predicting negative universally). To address this, we implement the PK-sampling strategy (\cite{hermans2017defense}), which constructs balanced epoch data by sampling an identical quantity of positive (Q-LERG) and negative (non-Q-LERG) samples per training iteration.
Specifically, each training epoch comprises a balanced subset formed by all positive samples and an equal-sized randomly sampled subset of negative samples. Upon completing an epoch, the negative samples are re-sampled to form a new balanced subset for the next epoch. This process repeats for 50 epochs, ensuring the model 1) to be exposed to most available
data while 2) to be able to mitigate the bias induced by class imbalance.

% {\bf \textit{Data Augmentation.}} 
Data augmentation is a widely adopted technique in machine learning to improve model generalization and robustness. In our training pipeline, we employed the following augmentation strategies:

\begin{itemize}
    \item Color Jitter: With a 50\% probability, we applied random perturbations to image brightness and contrast to simulate varying lighting conditions. This enhanced color diversity and mitigated overfitting to specific color distributions.
    \item Random Flip: Each image had a 50\% chance of being flipped either horizontally or vertically, encouraging the model to learn flip-invariant features.
    \item Random Rotation: Each image was randomly rotated with a 50\% probability, introducing rotational invariance into the learned representations.
    \item Random Erasing: With a 50\% probability, a randomly selected region of the image was erased. This encouraged the model to rely on holistic spatial features rather than overfitting to specific discriminative regions~(\cite{zhong2020random}).

\end{itemize}

% {\bf \textit{Model Inference.}} 
After training the model, for a sample $x$, the trained model can provide $\hat{y}$ to indicate the probability of $x$ belonging to a radio galaxy. We set a threshold $th$ to transfer the predicted continuous probability 
$\hat{y}$ into discrete classes as follows:
\begin{equation}
c =  \left\{
    \begin{aligned}
      &1 \quad &&\text{if } \hat{y} \ge th, \\
      &0 \quad &&\text{if } \hat{y} < th,
    \end{aligned}
  \right.
\end{equation}
where $c$ is the final predicted class and $th$ is a threshold. In the next section, we will show the choice of $th$ and the classification result.

\subsection{AI classification of RFE and RFI galaxies}

\begin{figure}
\includegraphics[width=\linewidth,keepaspectratio]{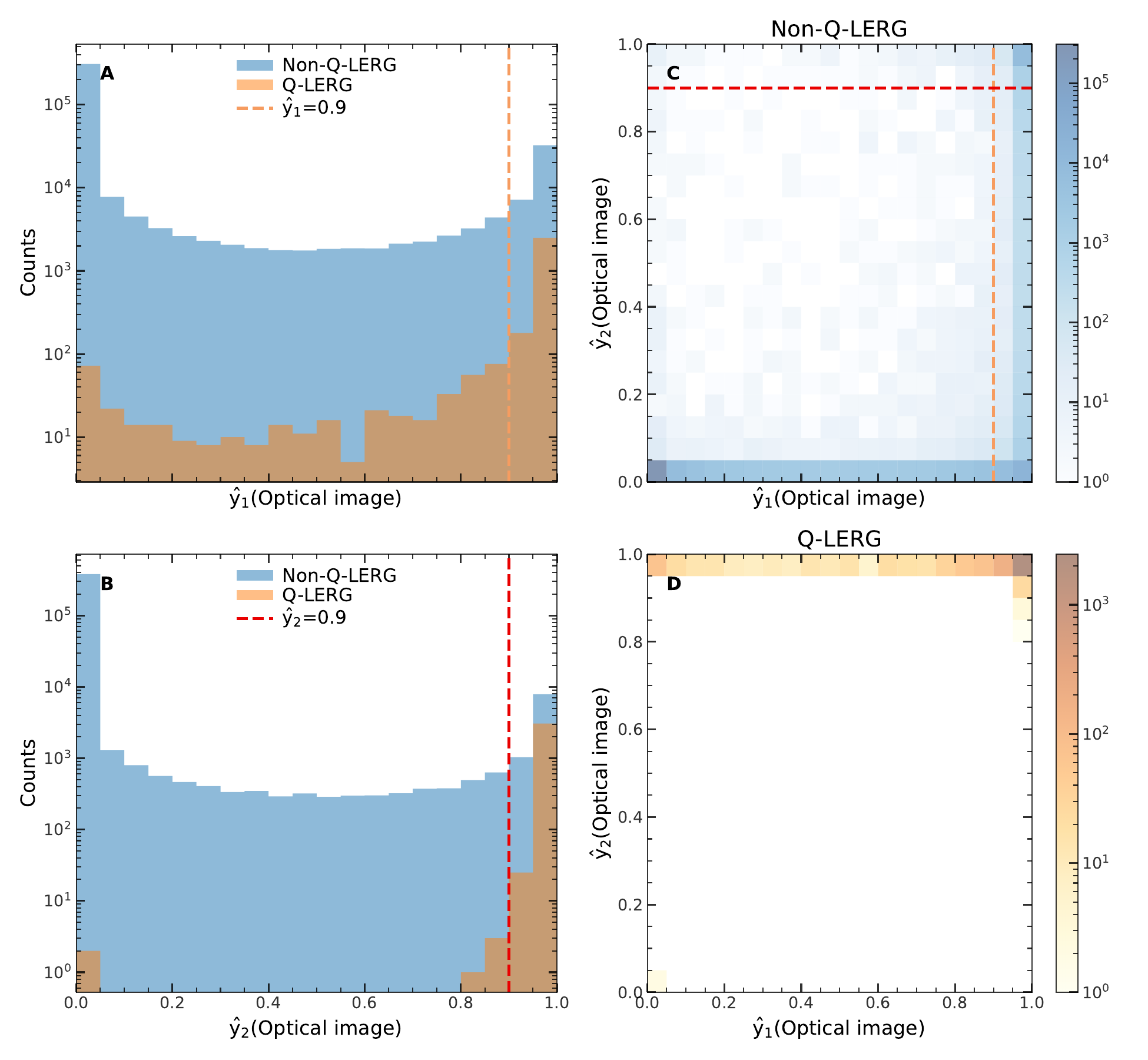}
\centering
\caption{{\bf Distributions of $\hat{y}_1$ and $\hat{y}_2$.} 
\textbf{A, B}: the number count histograms of $\hat{y}_1$ and $\hat{y}_2$ for Q-LERGs(orange) and non-Q-LERGs (blue). The dashed lines show the threshold of 0.9.
\textbf{C, D}: the 2D histograms of galaxy number. 
The parameter $\hat{y}_1$ and $\hat{y}_2$ are calculated by our AI technique based on the optical SDSS image with and without data augmentation, respectively. The parameters are used to evaluate the similarity between the images of unclassified galaxies and RFE-on galaxies. Large $\hat{y}$ means that the galaxies are similar to the RFE-on galaxies.}
\label{fig:p}
\end{figure}

\begin{figure*}
\includegraphics[width=0.8\linewidth, keepaspectratio]{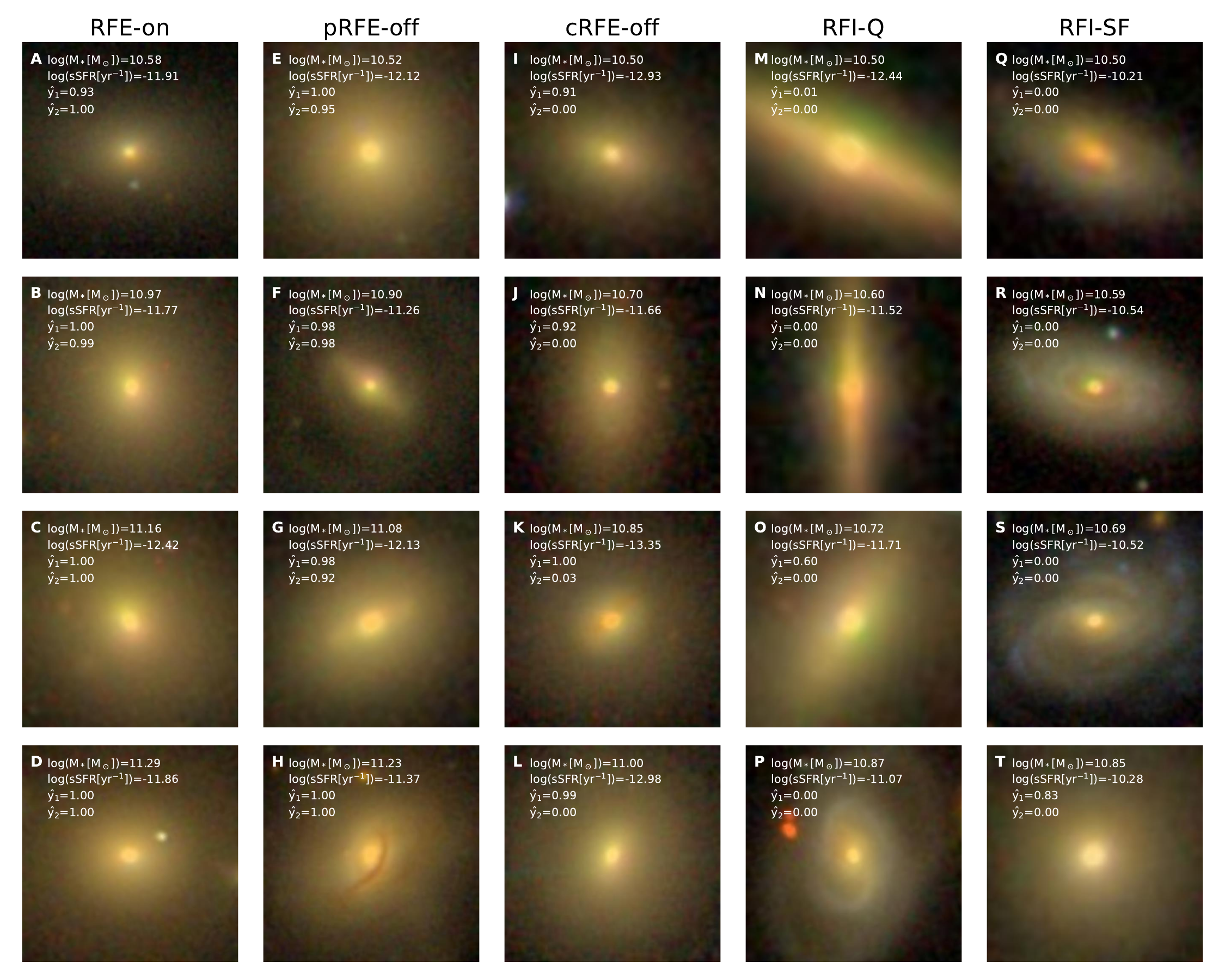}
\centering
\caption{{\bf Optical images for different types of galaxies.} 
\textbf{A to D}: the randomly selected images for RFE-on galaxies in four mass bins. \textbf{E to H}: pRFE galaxies. \textbf{I to L}: cRFE galaxies. \textbf{M to P}: RFI-Q galaxies. \textbf{Q to T}: RFI-SF galaxies.
The shown galaxies are selected from those at $z<0.04$ and with $\log M_\star/M_\odot>10.5$. The stellar masses, sSFR, $\hat{y}_1$ and $\hat{y}_2$ are shown in each panel.}\label{fig:gimage}
\end{figure*}

We applied our AI technique to the optical images to compute $\hat{y}_1$ (with data augmentation) and $\hat{y}_2$ (without data augmentation) for each galaxy. 
Fig.~\ref{fig:p} illustrates the distributions of $\hat{y}_1$ and $\hat{y}_2$ for Q-LERGs and non-Q-LERGs, respectively. 
Most Q-LERGs have both $\hat{y}_1$ and $\hat{y}_2$ close to 1, indicating that they share the same features. Therefore, all Q-LERGs are treated as RFE-on galaxies.
The $\hat{y}_1$ and $\hat{y}_2$ distributions for non-Q-LERGs have two distinct peaks at $0$ and $1$, respectively. This suggests that part of the non-Q-LERGs are very similar to Q-LERGs, while others are markedly different. 
Considering that most Q-LERGs exhibit $\hat{y}_1$ and $\hat{y}_2$ values exceeding 0.9, we opted to set $th=0.9$ as the threshold to identify RFE-off galaxies from non-Q-LERGs. Our tests reveal that the variation in $th$ from 0.2 to 0.9 has a negligible impact on our results. 
This is anticipated because $\hat{y}_1$ and $\hat{y}_2$ for most galaxies are close to either one or zero. 
Divergences between $\hat{y}_1$ and $\hat{y}_2$ occasionally occur (Fig.~\ref{fig:p}), highlighting the difficulty in classifying RFE and RFI galaxies solely based on noise labels. 
Consequently, we implemented a two-tier classification process. The results are listed in Table~\ref{tab:AIresult}.
Initially, we used $\hat{y}_1=0.9$ to divide non-Q-LERGs into two categories. Non-Q-LERGs with $\hat{y}_1<0.9$ are designated as the RFI sample, while those with $\hat{y}_1\ge0.9$ form the RFE-off sample with high completeness (cRFE-off). A total of 918 star-forming galaxies (0.4\% of the total population) have $\hat{y}_1\ge0.9$, but are actually classified as RFI galaxies.
Subsequently, a pure RFE-off (pRFE-off) sample was derived from the cRFE-off sample by imposing an additional requirement of $\hat{y}_2\ge0.9$. 
Thus, the pRFE-off sample constitutes a subset of the cRFE-off sample.
The pRFE sample is derived by merging the RFE-on and pRFE-off samples, whereas the cRFE sample is formed by combining the RFE-on and cRFE-off samples.
The galaxy numbers of these samples are listed in Table~\ref{tab_sample}.

Images randomly selected from RFE-on, pRFE-off, cRFE-off, RFI-Q, or RFI-SF samples at $z<0.04$ and with $\log (M_\star/M_\odot)>10.5$ are shown in Fig.~\ref{fig:gimage}.
RFE galaxies appear red, smooth, and spherical. RFI-Q galaxies tend to be slightly bluer and more elongated compared to RFE galaxies. RFI-SF galaxies are usually blue, and some of them exhibit spiral arms. A more detailed discussion will be presented in Section~\ref{sec:gal props}.

\subsection{Purity and Completeness of RFE-off samples}

\begin{figure*}
\includegraphics[width=0.85\linewidth,keepaspectratio]{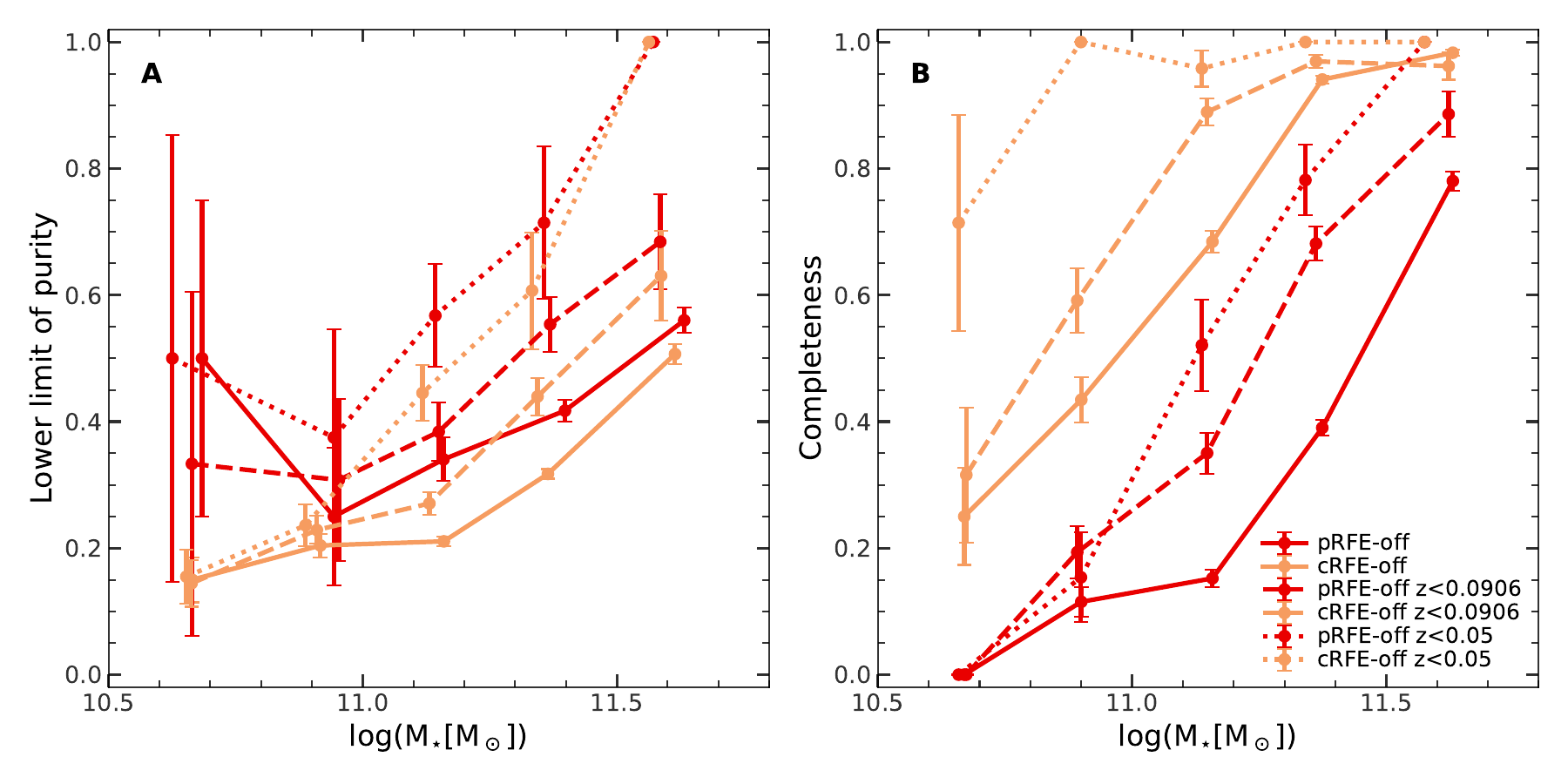}
\centering
\caption{{\bf Completeness and purity.} 
\textbf{A}: the lower limit of purity as a function of $M_\star$. The red lines show the results for pRFE-off sample, and the orange lines show the results for cRFE-off sample. The solid lines show the results using all galaxies with $z<0.2$, dotted line for galaxies with $z<0.0906$ and dashed lines for galaxies with $z<0.05$.
\textbf{B}: the completeness as a function of $M_\star$. The completeness, estimated using a standard 90/10 train-test split of the data. There is only 34 Q-LERGs (at $z<0.2$) in the lowest $M_\star$ bin, leading to large uncertainties in completeness.
Purity refers to the fraction of selected RFE-off galaxies that are observationally confirmed in LOTSS DR2. Error bars represent Poisson 1$\sigma$ uncertainties.
}\label{fig:AI_performance}
\end{figure*}

The performance of a classification method is usually assessed by two metrics, completeness and purity. 
In this section, we evaluate the two metrics for RFE-off galaxies.
Completeness refers to the proportion of intrinsic RFE-off galaxies that are correctly classified, while 
purity refers to the proportion of galaxies classified as RFE-off galaxies that are intrinsically RFE-off galaxies. 
A standard approach to assess the two parameters is to apply the method to a test sample, within which galaxies have been accurately classified by an independent and reliable method. 
However,  there are no well-defined RFE-off sample in the literature, because RFE-off and RFI-Q galaxies occupy the same region in the AGN-SF diagram (Fig.~\ref{fig:AGNSF}).
Therefore, we need an alternative method.

To evaluate the completeness, we randomly selected 90\% of Q-LERGs (RFE-on) for training purposes. 
We then applied the trained AI to the rest 10\% Q-LERGs which are treated as intrinsic RFE-off galaxies in the test. 
The approach is reasonable because RFE-off galaxies have optical images similar to those of RFE-on galaxies. 
We conducted 10 such tests with randomly selected test galaxies. The completeness is calculated as $C=N_{\rm s,QL}/N_{\rm t, QL}$, where $N_{\rm t,QL}$ is the number of test Q-LERGs and $N_{\rm s,QL}$ is the number of test Q-LERGs classified as RFE galaxies.

\begin{figure}
\includegraphics[width=\linewidth,keepaspectratio]{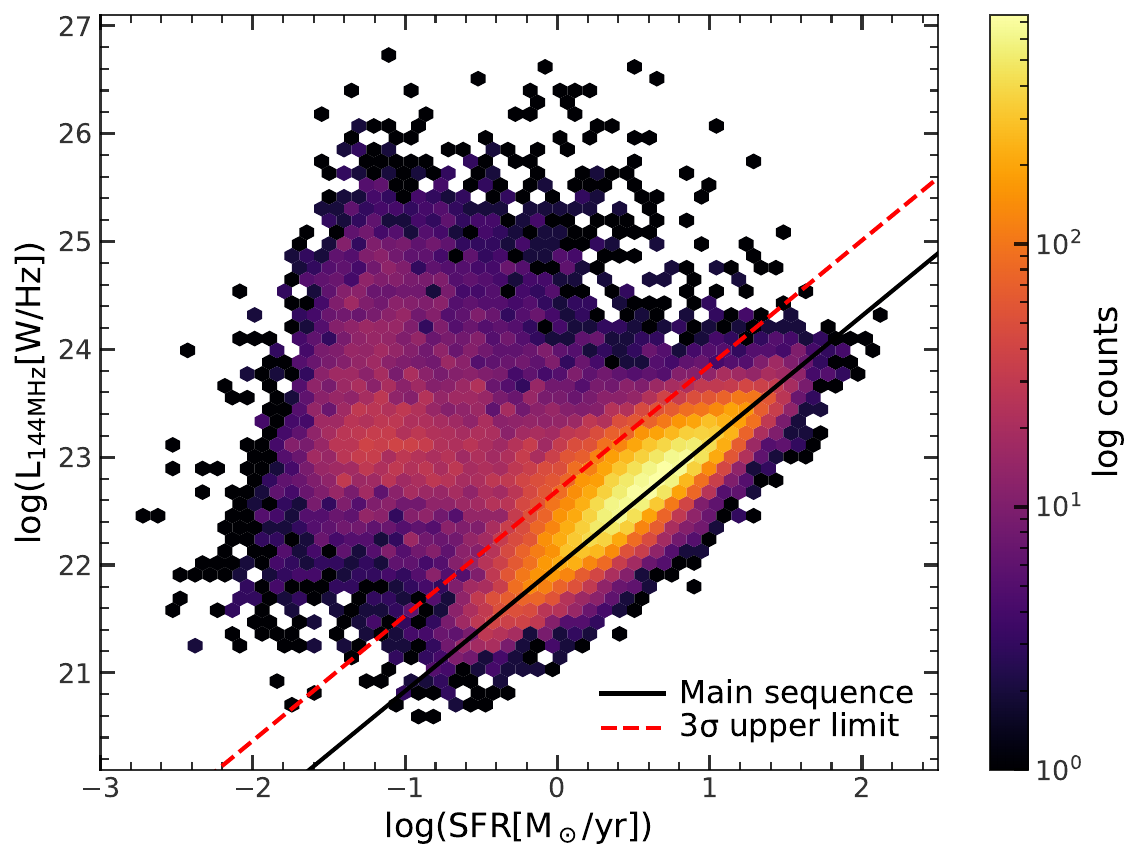}
\centering
\caption{{\bf Distribution of LOTSS DR2 galaxies in L-SFR plane.} The background shows a 2D histogram (hexbin) of galaxy number as indicated by the color bar. The solid black line indicates the star-forming main sequence ($L_{\rm{SFR,144MHz}}$-SFR relation, \cite{Jin2025}), while the dashed red line shows the relation plus 3$\sigma$, where $\sigma=0.2$ dex is the scatter of the relation.
}\label{fig:L_SFR}
\end{figure}

The purity of RFE-off samples is estimated using a deeper radio survey, the LOFAR Two-Metre Sky Survey DR2 (LOTSS DR2 \cite{Hardcastle2023}), observed at a central frequency of 144 MHz.  
Fig.~\ref{fig:L_SFR} shows the radio luminosity, $L_{\rm 144MHz}$, versus SFR for the detected galaxies. There are two populations. 
The first population has $L_{\rm 144MHz}$ closely correlated with SFR, following well the $L_{\rm SFR,144MHz}$-SFR relation for star-forming galaxies derived by \cite{Jin2025}, indicative of the star-formation origin of radio luminosity. 
The other population lies above the star-forming main sequence. The excess radio luminosity originates from radio AGNs. Following  \cite{Jin2025}, 
we selected radio-excess galaxies using
\begin{equation}
    \log(L_{\text{144MHz}}[\text{W/Hz}]) \ge L_{\rm SFR,144MHz}+3\sigma_{\rm L}
\end{equation}
where $L_{\rm SFR,144MHz}$ is the $L_{\rm 144MHz}$-SFR relation for star-forming galaxies (Fig.~\ref{fig:L_SFR}) 
\begin{equation}
     L_{\rm SFR,144MHz}= 1.16\log(\text{SFR}[M_{\odot}/\text{yr}]) + 21.99
\end{equation}
and $\sigma_{\rm L}=0.23$ is the scatter of the relation (\cite{Jin2025}). 

We assume that a galaxy exhibiting radio excess and yet classified as RFE-off represents an intrinsic RFE-off galaxy, undetectable by the NVSS/FIRST survey but discernible through the LOTSS survey. This galaxy is likely experiencing
low-level activity within its duty cycle. The low-limit of the purity is defined by $P_{\rm ll}=N_{\rm off, re}/N_{\rm off}$, 
where $N_{\rm off}$ corresponds to the count of RFE-off galaxies within the selected sky region, and $N_{\rm off, re}$ denotes the number of RFE-off galaxies manifesting radio excess at 144MHz. $P_{\rm ll}$ serves as a low-limit due to the fact that some intrinsic RFE-off galaxies possess radio activity too faint for LOTSS detection. 
Here, we emphasize that achieving $P_{\rm ll}$ =100\% is unrealistic even for an intrinsically “RFE” sample. This is because RFE-off systems exhibit little or no ongoing radio activity and would remain undetected even in the deepest available radio surveys. 
%Moreover, the fact that our $P_{\rm ll}$ does not approach unity also suggests that the morphologies of RFE-off and RFE-on galaxies are not dramatically different; if their structural differences were sufficiently strong and unambiguous, one would expect the classification probability to saturate much closer to 100\%. 

\begin{figure*}
	\includegraphics[width=0.8\linewidth,keepaspectratio]{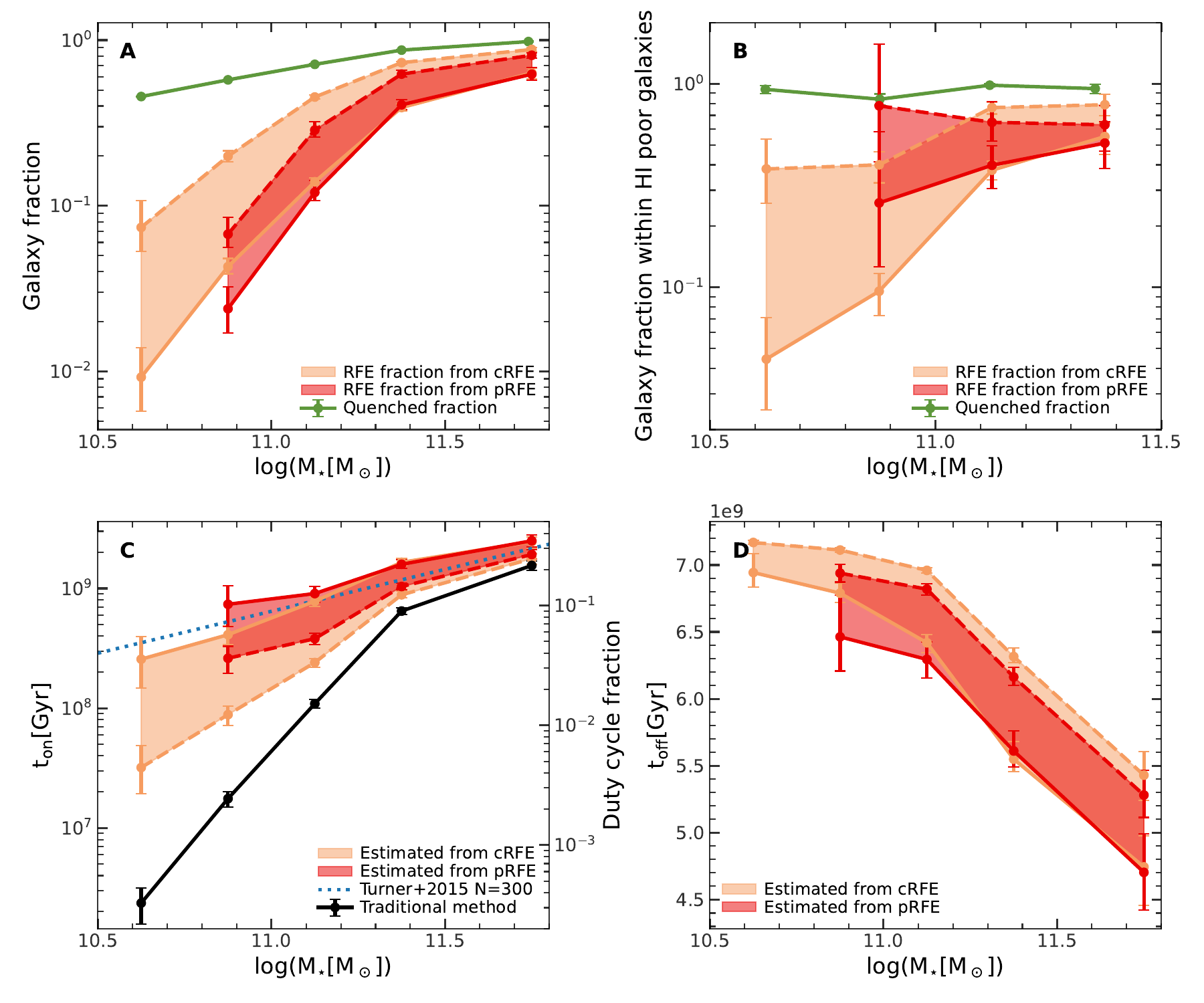} 
    \centering
	\caption{\textbf{Galaxy fractions and duty cycle.} \textbf{A}: fractions of quenched galaxies (green), 
RFE galaxies estimated from the pRFE sample (red) and the cRFE sample (orange) as functions of $M_\star$.  
\textbf{B}: quenched and RFE fractions in the gas-poor population of galaxies ($\log M_{\rm HI}/M_\star<-1.5$).
\textbf{C}: the on-duty duration time ($t_{\rm on}$, left vertical axis) and duty cycle fraction (right axis) for radio galaxies with $L_{\rm 1.4GHz}>10^{23}\rm W~Hz^{-1}$.
\textbf{D}: the off-duty duration time ($t_{\rm off}$) as functions of $M_\star$.
Error bars represent the 16th–84th percentiles of 150 bootstrap samples. 
The dashed line(red or orange) is obtained by assuming 100\% purity and the solid line corresponds to the lower limit of the purity.
}
	\label{fig:gf}
\end{figure*}

The completeness of the whole cRFE-off and pRFE-off samples is 85.0\% and 40.2\%, respectively. 
The average $P_{\rm ll}$ for pRFE-off galaxies stands at 46\%, significantly surpassing that of cRFE-off galaxies at 29.5\%. 
Fig.~\ref{fig:AI_performance} shows C and $P_{\rm ll}$ as functions of $M_\star$. The two metrics for both cRFE and pRFE increase strongly with increasing $M_\star$, indicative of the low efficiency of the method at the low-mass end. 
We also assessed the two metrics for galaxies at $z<0.0906$, where radio galaxies with $L_{\rm 1.4GHz}>10^{23}\rm~WHz^{-1}$ are complete and galaxies with $\log M_\star/M_\odot>10.5$ are also nearly complete in SDSS. 
This sample can be used to estimate the RFE fraction and duty cycle(Fig.~\ref{fig:gf}).  
The mean $C$ for pRFE-off and cRFE-off samples are 52\% and 87\%, respectively. The mean $P_{\rm ll}$ for pRFE-off and cRFE-off samples are 49\% and 30\%, respectively.
We also assessed the two metrics for galaxies at $z<0.05$,
which is the redshift range for xGASS galaxies. 
The mean $C$ for pRFE-off and cRFE-off samples are 55\% and 97\%, respectively. The mean $P_{\rm ll}$ for pRFE-off and cRFE-off samples are 60\% and 32\%, respectively.
The two metrics can be used to estimate the RFE fraction in gas-poor galaxies (Fig.~\ref{fig:gf}).
As can be seen, both metrics improve as the redshift decreases. One possible reason for the improvement is that galaxies at lower redshift have images of better quality. 

%__________________________________________________________________

\section{Fraction of RFE galaxies and duty cycle of radio activity}
\label{sec:galaxy fractions}

\subsection{RFE galaxy fraction}

% For a given RFE (cRFE or pRFE) sample, we calculated the fraction of RFE galaxies as
In order to assess the impact of radio feedback across the quenched galaxies, we calculated the fraction of RFE galaxies,
\begin{equation}
    f_{\rm RFE}=\frac{N_{\rm off}*P/C+N_{\rm on}}{N_{\rm all}}\\,
\end{equation}
where $N_{\rm all}$ is the number of galaxies in the total sample, $N_{\rm on}$ and $N_{\rm off}$ are the numbers of RFE-on and RFE-off galaxies in the RFE sample, respectively, and $P$ and $C$ are the purity and completeness of the RFE-off sample.
The fraction of quenched galaxies is $f_{\rm q}=N_{\rm q}/N_{\rm all}$, where $N_{\rm q}$ is the number of quenched galaxies. 
Since it is difficult to measure a precise value of $P$, we present the results with $P=P_{\rm ll}$ and $P=1$ in Fig~\ref{fig:gf}A. Note that the RFE fraction and quenched fraction are calculated using galaxy sample, completeness and purity with $z<0.0906$.

Fig~\ref{fig:gf}A shows the fraction of RFE galaxies as a function of $M_\star$, 
using cRFE and pRFE samples, respectively. Both cRFE and pRFE samples yield similar results, demonstrating the robustness of our AI method.
The RFE fraction increases from (0.9 - 7) percent at $\log M_\star/M_\odot=10.6$ to 
(5 - 30) percent at $\log M_\star/M_\odot=11$. In the most massive bin, nearly all galaxies are RFEs.
To examine the role of radio AGNs in maintaining the quiescence of galaxies, we compared the RFE fraction with the fraction of quenched galaxies. 
In the most massive bin, the RFE and quenched fractions are comparable, suggesting that radio feedback can help maintain the quiescence of these galaxies. 
At $\log M_\star/M_\odot \sim11.5$, the difference between the two fractions 
becomes visible. 
While at $\log M_\star/M_\odot \sim11$, the quenched fraction, at about 62\%, is much higher than the RFE fraction (with a median value of $\sim17$\%). 
It is evident that the quenched fraction is significantly larger than the RFE fraction for  $\log M_\star/M_\odot<11.5$, indicating that radio feedback cannot explain the quiescence of most galaxies with $\log M_\star/M_\odot <11.5$, and that there must be other channels to keep massive galaxies quiescent. 

\subsection{Duty cycle}

%\begin{figure*}
%\includegraphics[width=0.85\linewidth, keepaspectratio]{Figs/figex_age_profile.pdf}
%\centering
%\caption{{\bf Stellar age radial profile from MaNGA} 
%The x-axis shows radius normalized by the effective radius $R_{\rm{50}}$ , and the y-axis represents the mass-weighted stellar age in Gyr. Shaded regions show the 16th–84th percentile range. {\color{red} seems not necessary}}\label{fig:age_profile}
%\end{figure*}

Radio galaxies are known to experience cycles of activity and silence.
We calculated the duty cycle fraction as
\begin{equation}
    f_{\rm dc}=\frac{N(>L_{\rm th})}{N_{\rm on}+N_{\rm off}*P/C}\\,
\end{equation}
where $N(>L_{\rm th})$ is the number of RFE galaxies with radio luminosity above $L_{\rm th}$, $N_{\rm on}$ is the number of RFE-on galaxies, $N_{\rm off}$ is the number of RFE-off galaxies, and $P$ and $C$ are the purity and completeness of the sample. Note that the denominator is actually the intrinsic number of RFE galaxies.
Since most RFE-on galaxies have $L_{\rm 1.4GHz}>10^{23}\rm W~Hz^{-1}$, we set $L_{\rm th}=10^{23}\rm W~Hz^{-1}$. Given the flux limit of 5 mJy, radio galaxies above this threshold are complete at $z<0.0906$. The completeness and purity in this redshift range are shown in Fig.~\ref{fig:AI_performance}. 
For comparison, we also estimated the duty cycle fraction using a traditional approach, defined as the fraction of Q-LERGs ($L_{\rm 1.4GHz}>10^{23}\rm W~Hz^{-1}$) among all galaxies.  Galaxies with $\log M_\star/M_\odot>10.5$ are also complete in this redshift range, so we only show the results for  $\log M_\star/M_\odot>10.5$ (Fig.~\ref{fig:gf}C). 
Similar fractions were obtained in previous studies(\cite{Best2012, Sabater2019, Kondapally2024}), however, using all galaxies rather than RFE galaxies.
These results cannot be interpreted as the duty cycle of maintenance-mode radio AGNs. 
This is because RFI galaxies, which are not regulated by radio feedback, are included in the calculation. 
We emphasize that the duty cycle fraction is impossible to estimate if we do not have a reasonable estimation for the number of RFE galaxies. 
As can be seen, the duty cycle increases from $\sim1\%$ at $\log M_\star/M_\odot=10.6$ to $\sim7\%$ at $\log M_\star/M_\odot=11$, and then to $\sim30\%$ at $\log M_\star/M_\odot=11.8$. 
These are much larger than the duty-cycle fraction obtained using the traditional method.

To convert the duty cycle fraction into physical time durations, we assume a total RFE timescale of $t_{\rm{RFE}} = 7.2~\rm{Gyr}$, motivated by the stellar population properties of RFE galaxies with $\log M_\star/M_\odot>10.5$ in the MaNGA survey (\cite{Bundy2015, Sanchez2022}). 
We found that the stellar age radial profiles of these RFE galaxies show little dependence on radius, with a median mass-weighted stellar age of approximately $7.2\pm 1.1$ Gyr. We then adopt $t_{\rm{RFE}}=7.2~\rm{Gyr}$ as a representative time scale for the maintenance of quenching by radio feedback. 
The on-duty duration is then calculated as
\begin{equation}
    t_{\rm{on}} = f_{\rm{dc}} t_{\rm{RFE}}\\.
\end{equation}
The corresponding off-duty time is 
$t_{\rm{off}} = t_{\rm{RFE}} - t_{\rm{on}}$. 
Fig.~\ref{fig:gf}C and D show $t_{\rm on}$ and $t_{\rm off}$ as functions of $M_\star$. 
\cite{Turner2015} derived the duration of a single radio AGN activity as a function of $M_\star$, %based on a volume-limit radio AGN sample,
\begin{equation}
    t_{\rm{on, single}} = 4.8~{\rm Myr}~f^{-1/2} (M_\star/{\rm M}_\odot)^{0.7}
\end{equation}
where $f$ is a scaling factor in their jet power–luminosity relation, for which we adopt $f=5$. 
Interestingly, the slope of the $t_{\rm on, single}-M_\star$ relation is similar to our $t_{\rm on}-M_\star$ relation.
We scaled the $t_{\rm on, single}-M_\star$ relation by 300 and found that it matches our result very well (blue dashed line in Fig.~\ref{fig:gf}). This suggests that on average an RFE galaxy has experienced around 300 cycles.

%__________________________________________________________________

\section{Properties of RFE and RFI Galaxies}
\label{sec:gal props}

\begin{figure*}
	\includegraphics[width=\linewidth,keepaspectratio]{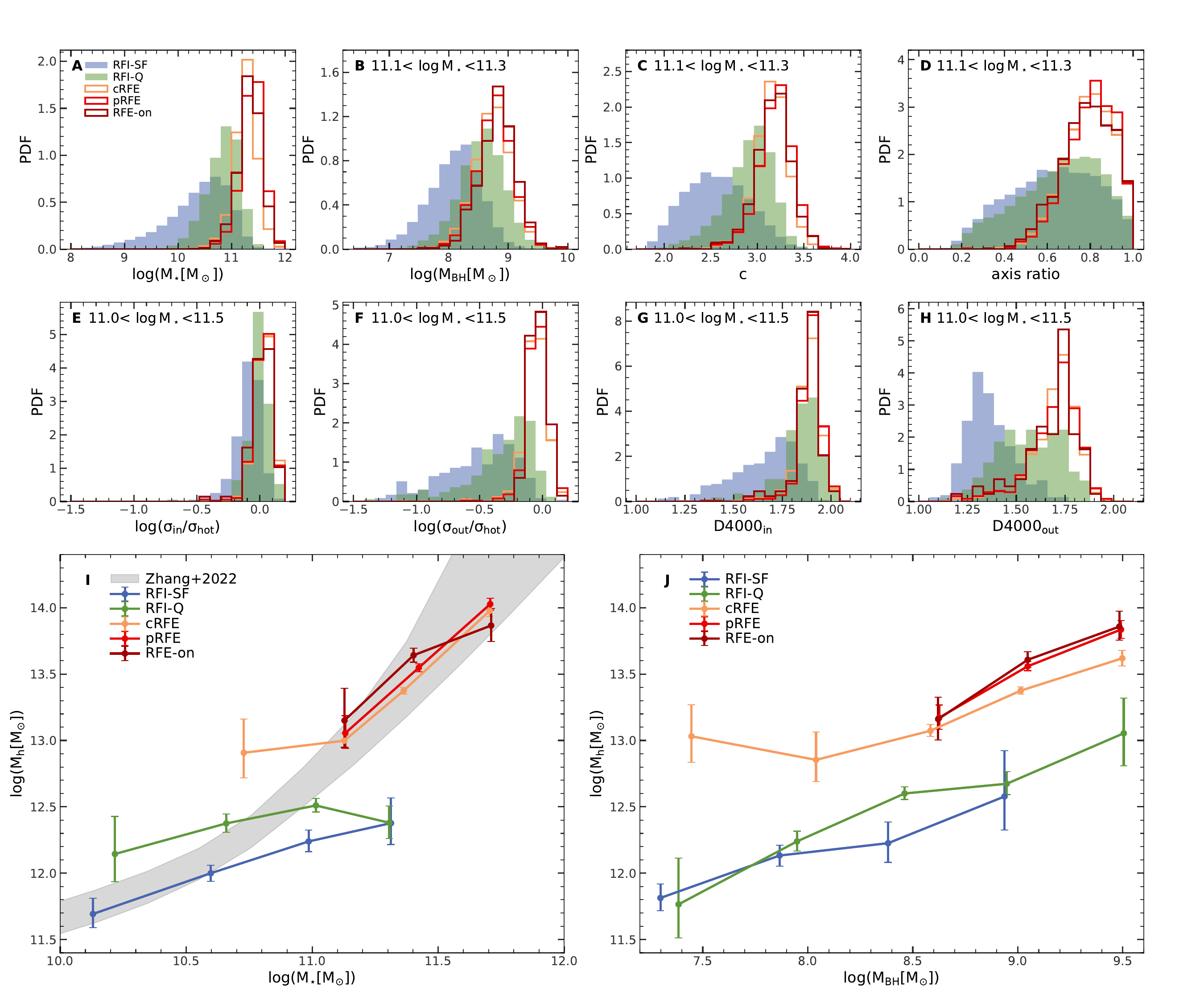} 
	\caption{\textbf{Galaxy and halo properties of RFE and RFI galaxies.}
		\textbf{A}: the probability distributions of $M_\star$ for RFE-on (wine), pRFE (red), cRFE (orange), RFI-Q (green) and RFI-SF (blue) galaxies. 
\textbf{B to D}: the distributions of $M_{\rm BH}$, concentration and axis ratio for the five types of galaxies with $11.1<\log M_\star/M_\odot<11.3$. 
\textbf{E to H}: the distributions of spatially-resolved galaxy properties for the five types of galaxies with $11.0<\log M_\star/M_\odot<11.5$, including the stellar velocity dispersion in the inner ($\sigma_{\rm in}$) 
and outer regions ($\sigma_{\rm out}$) scaled with the velocity dispersion ($\sigma_{\rm hot}$) of a 
dynamically hot system at the same $M_\star$ (\cite{HongH2023}), and the 4000\AA\ break in the inner and outer regions. 
A system is dynamically hot if its velocity dispersion is comparable to $\sigma_{\rm hot}$,  
otherwise it is dynamically cold, usually dominated by rotational motion. 
Note that results in other $M_\star$ bins are similar but not plotted.
\textbf{I, J}: the $M_{*}$-$M_{\rm h}$ and $M_{\rm BH}$-$M_{\rm h}$ relations, where halo masses are estimated 
using weak gravitational lensing technique. The shaded region represents the combination of several empirical 
$M_{*}$-$M_{\rm h}$ relations (\cite{ZhangZ2022}). We only show results for bins with errors less than 0.5 dex 
and with galaxy numbers above 100. Error bars represent the 16th–84th percentiles of 150 bootstrap samples. 
In general, RFE galaxies are different from RFI galaxies in dynamics, stellar population, morphology and halo mass. 
Both pRFE and cRFE galaxies exhibit similar properties to RFE-on galaxies. %{\color{red}separate this figure into two figures.}
}
	\label{fig:mps}
\end{figure*}

\subsection{Galaxies properties}

Fig. \ref{fig:mps} shows the probability distributions of various properties for RFE and RFE-on galaxies. 
Since the fraction of RFE-on galaxies in both RFE samples remains consistently low 
(Fig.~\ref{fig:gf}C), the results for RFE galaxies are dominated by RFE-offs.
As can be seen, the differences between RFE and RFE-on galaxies are very small in all the properties in consideration, including stellar mass, black hole mass, stellar population, morphology, and dynamical properties. The differences between cRFE and RFE-on are slightly larger, possibly due to larger contamination in the cRFE sample. These results suggest that our AI method reliably achieves its purpose, that is, to select galaxies similar to RFE-on galaxies.

We then compare RFE galaxies and RFI galaxies, which are expected to evolve in different paths.
For the RFI population, we focused on RFI-Q galaxies, which have a star-formation status similar to that of RFE galaxies. 
The most notable distinction lies in the $M_\star$ distribution. RFE galaxies tend to be more massive compared to RFI-Qs. 
For a fixed $M_\star$, RFE galaxies also possess more massive black holes, have higher concentration and larger axis ratio than RFI-Q galaxies. 
The distribution of axis ratio for RFI-Qs, similar to that for RFI-SFs, is much broader than that of RFE galaxies. 
Interestingly, the distributions for RFI-Q and RFI-SF galaxies are very similar to that for disk galaxies as shown in \cite{ZhangY2015}. This means that these two types of galaxies have remarkable disk components. The galaxies with small axis ratio are very likely edge-on galaxies. In contrast, none of RFE galaxies shows small axis ratio. Therefore, the disk component in RFE galaxies should be very weak or even absent.

Fig.~\ref{fig:mps} also shows the distributions of four parameters, including $\sigma_{\rm in}$, $\sigma_{\rm out}$, D4000$^{\rm in}$ and D4000$^{\rm out}$, for the MaNGA galaxies with $11.0<\log M_\star/M_\odot<11.5$. Note that the two dynamical parameters are scaled with $\sigma_{\rm hot}(M_\star)$, which is the scaling relation of dynamical hotness and is taken from our previous study (\cite{HongH2023}). A region is considered dynamically hot when its velocity dispersion is comparable to $\sigma_{\rm hot}(M_\star)$.
The central regions of both RFE and RFI-Q galaxies show similar stellar populations and dynamical properties, as indicated by the 4000\AA\ break (D4000) and velocity dispersion ($\sigma$), respectively. 
In the outskirts, however, the stellar populations of RFE galaxies generally appear older than those of RFI-Q galaxies. 
The outskirts of RFE galaxies are dynamically hot, similar to their 
central regions, whereas the outskirts of RFIs are dynamically much colder. 
The similarity in the axis ratio and kinematics in the outskirts of both RFI-Q and RFI-SF galaxies 
suggests that the outskirts of RFI-Qs are primarily cold, rotation-supported disks. 
In contrast, the higher axis ratio and velocity dispersion indicate that RFE galaxies are nearly spherical
and dominated by random motion. 
Our finding aligns with previous studies (\cite{Heckman2014, Barisic2019, HongH2023}), although they did not differentiate between RFE-off and RFI galaxies. 
This result is consistent with the idea that the AI technique differentiates RFE and RFI galaxies primarily based on the outskirt image(see Appendix).

\subsection{Host halo masses}

%\begin{figure*}
%\includegraphics[width=\linewidth,keepaspectratio]{Figs/figex_SHMR.pdf}
%\centering
%\caption{{\bf Stellar mass-Halo mass relation and Black hole mass-Halo mass relation.} \textbf{A}: $M_\star$-$M_{\rm h}$ relation for different populations. The grey shaded region represents the empirical relation from \cite{ZhangZ2022}. \textbf{B}: $M_{\rm BH}$–$M_{\rm h}$ relation. The halo masses are estimated by using weak lensing technique. We only show the results for bins with error less than 0.5 dex and galaxy numbers above 100. The error bars represent the 16th–84th percentiles of 150 bootstrap samples.} \label{figex_SHMR}
%\end{figure*}

Earlier research suggests that halos hosting radio galaxies are approximately twice as massive as those hosting other galaxies of the same $M_\star$ (\cite{Mandelbaum2009}). Using weak gravitational lensing (\cite{Zhang_2019}),
we obtained the average halo mass ($M_{\rm h}$) for various populations of galaxies (see Figure~\ref{fig:mps}). Please see Appendix for the details of the weak lensing technique.
At given $M_\star$, the difference in $M_{\rm h}$ between RFE and RFI-Q 
galaxies is about 0.6 to 1 dex, much higher than previous estimates of about 0.3 dex (\cite{Mandelbaum2009}). 
This can be attributed to the proficiency of the AI classifier in distinguishing RFE-off from RFI-Q galaxies under the expectation that RFE-offs and RFE-ons have similar $M_{\rm h}$. 
These results suggest that the global efficiency to convert gas into stars, characterized by $M_\star/M_{\rm h}$, is about 4 to 10 times higher in RFI galaxies than in RFE galaxies. This is consistent with the assumption that AGN feedback is much more effective in RFE galaxies than in RFIs. 

It has been proposed recently that the $M_{\rm h}$-$M_{\rm BH}$ relation can provide insight into AGN feedback mechanisms (\cite{Shankar2020, LiQ2024}). 
We thus also present such relations based on our classification. 
At fixed $M_{\rm BH}$, the difference in $M_{\rm h}$ between the RFE and RFI-Q galaxies ranges from 0.6 to 0.8 dex. 
More interestingly, the RFE and RFI galaxies appear to reside in distinct halo populations, with a division at $M_h\sim10^{12.8}M_\odot$. 
This demonstrates that $M_{\rm h}$ plays an important role in triggering radio AGN feedback, potentially more important than $M_\star$ and $M_{\rm BH}$. 
One possibility is that the higher merger rate in more massive halos enhances the 
spin of supermassive black holes, thereby facilitating the launching of radio jets (\cite{Blandford1977, YangH2024}).
Alternatively, the increase in the amount of hot gas in more massive halos may promote AGN formation and feedback. 

The galaxy images used in AI learning are not expected to carry any direct information related to $M_{\rm h}$ since we only used the galaxy images within 3$R_{50}$. 
Therefore, the large difference in $M_{\rm h}$ between RFEs and RFIs thus testsifies to the robustness of our AI classification. 
At fixed $M_\star$, the halo and galaxy characteristics of both pRFE and RFE-on galaxies are very similar, consistent with our AI method's assumption that RFE-off and RFE-on galaxies are similar in galaxy and halo properties. 
The difference between cRFE and RFE-on galaxies is marginally larger, probably due to the elevated contamination in the cRFE sample.
Many studies developed methods to use galaxy properties, such as stellar mass, velocity dispersion and others, to infer halo mass(see e.g. \cite{ZhangZ2024}). Our work suggests that in order to derive a more accurate estimation of halo mass, one should consider galaxy properties that are used by the AI method to classify RFE and RFI galaxies.

%In Fig.~\ref{figex_SHMR}, we show $M_{\rm h}$ as function of $M_\star$ and $M_{\rm BH}$ for Q-LERGs and non-Q-LERGs. Consistent with previous study (\cite{Mandelbaum2009}), the difference $M_{\rm h}$ between the two populations at given $M_\star$ is about 0.3 dex. For comparison, we also present the results for pRFE-off, cRFE-off and RFI galaxies. Similarly to that in Fig.~\ref{fig:mps}, the difference between RFE and RFI galaxies is very large. Apparently, the non-Q-LERG sample comprises two distinct populations, RFI and RFE-off galaxies. RFE-off galaxies reside in massive halos, comparable to those of Q-LERGs, while the host halos of RFI galaxies are much smaller. 

%__________________________________________________________________

\section{HI in galaxy}
\label{sec:gas}

\begin{figure*}
	\centering
	\includegraphics[width=1.0\linewidth]{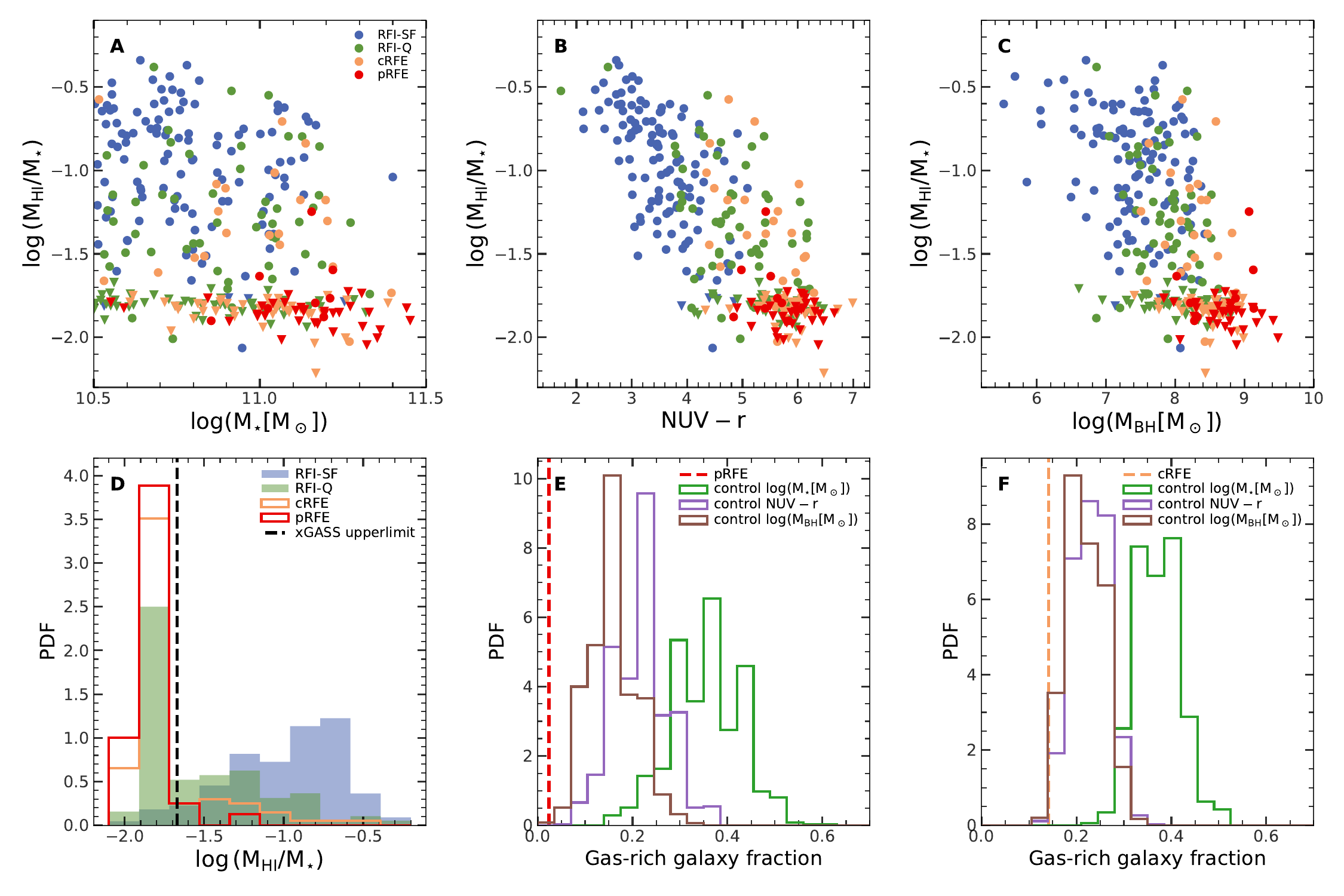} 
	\caption{\textbf{The HI-gas fraction.}
		\textbf{A to C}: the HI gas fraction ($\log M_{\rm HI}/M_\star$, obtained from the xGASS survey) as a function of various galaxy properties: stellar mass $M_\star$, NUV-r color, and black hole mass 
$M_{\rm{BH}}$. Triangle symbols show upper limits of $\log M_{\rm HI}/M_\star$. \textbf{D}: the histograms of 
$\log M_{\rm HI}/M_\star$ for xGASS galaxies. The dashed black vertical line shows the maximum value of the 
upper-limits. In panels \textbf{A to D}, red is for pRFE, orange for cRFE, green for RFI-Q and blue for RFI-SF.
\textbf{E}: the distribution of the gas-rich fraction (fraction of galaxies with $\log M_{\rm HI}/M_\star> -1.5$ in a sample). 
The vertical dashed line shows the result for the pRFE sample. The histograms show distributions of control samples
selected from the RFI-Q sample to match the pRFE sample in $M_{*}$ (green), NUV-r (purple) and $M_{\rm BH}$(brown), 
respectively. \textbf{F}: similar to \textbf{E} but showing results for cRFE galaxies. All the results are based on 
galaxies with $\log M_\star/M_\odot>10.5$. The AI-selected RFE galaxies have much less HI gas than RFI-Q galaxies 
at given $M_\star$, $M_{\rm BH}$ or NUV-r color. Since most of the gas-poor galaxies are RFE galaxies after 
the correction of misidentification (Fig.~\ref{fig:gf}), the real gas-rich fraction of RFI-Q galaxies is 
much larger than that shown here. Note that there are only 6 RFE-on galaxies, we do not show them separately.}
	\label{fig:HI}
\end{figure*}

The classification of RFE and RFI galaxies provides a unique opportunity to study the impact of radio feedback on galaxies. 
Should radio AGN feedback be capable of effectively heating the circumgalactic medium (CGM), there must be a large difference in the cold gas component between RFE and RFI galaxies. 
To test this explicitly, we compared the HI content between these two populations of galaxies. We obtained the HI gas fraction $M_{\rm HI}/M_\star$ for our galaxy samples provided by the xGASS survey (\cite{Catinella2018}). There are 6 RFE-on galaxies, 42 pRFE galaxies, 106 cRFE galaxies, 101 RFI-Q galaxies and 116 RFI-SF galaxies with available HI gas fraction.
Fig.~\ref{fig:HI}A shows the HI gas fraction of these galaxies.
Given that most RFE galaxies have $\log M_\star/M_\odot>10.5$, our analysis was restricted to galaxies above this mass limit.
The HI gas fraction increases progressively as the sample changes across the sequence pRFE, cRFE, RFI-Q, and RFI-SF. Most pRFE galaxies are deficient of HI gas, with $\log M_{\rm HI}/M_\star<-1.7$. 
Of the 42 pRFE galaxies with available xGASS data, only one has $\log M_{\rm HI}/M_\star>-1.5$ (thus gas-rich). 
Among the 106 cRFE galaxies, 15 are classified as gas-rich, and the 
larger gas-rich fraction is probably due to larger contamination in this sample. 
Within the subset of 6 RFE-on galaxies, none is gas-rich. In contrast, 38\% of RFI-Q galaxies and 90\% of RFI-SF galaxies are rich in HI gas. 
The gas-rich threshold is set to be $\log M_{\rm HI}/M_\star=-1.5$, 
because the distribution of star-forming galaxies quickly declines around this value(Fig.~\ref{fig:HI}D).

To mitigate the potential dependence on $M_\star$, we randomly chose one RFI-Q galaxy that matches each pRFE or cRFE galaxy in $M_\star$. For each galaxy in pRFE (cRFE) sample, a matched galaxy is randomly chosen from the RFI-Q sample with a mass difference of $<0.1$ dex. 
Note that if there is no matched galaxy within the tolerance, a galaxy with the smallest difference in the property considered is selected.
A total of 1000 control samples are constructed of RFI-Q galaxies in this way. 
For each control sample, we computed the fraction of gas-rich galaxies. 
As shown in Fig.~\ref{fig:HI}, the median value of the gas-rich fraction is 36\% (37\%) for control samples constructed to match pRFE (cRFE). 
Even the lowest values obtained from the control samples for pRFE and cRFE, 14\% and 23\%, respectively. They are much higher than the values for the pRFE and cRFE samples themselves. 
Several studies have found that the HI gas fraction has the strongest correlation with the NUV-r color (\cite{Catinella2018}), $M_{\rm BH}$ (\cite{WangT2024Natur}) and the linear combination of NUV$-$r and the stellar surface density (\cite{Catinella2013}). 
%As shown in Figure~\ref{fig:HI}, the correlation with the color and $M_{\rm BH}$ is indeed stronger than that with $M_\star$. 
We constructed control samples from RFI-Q by matching the NUV-r color with pRFE or cRFE galaxies. The tolerance is $\rm\Delta(NUV-r)<0.3$.
The samples controlled in NUV-r have gas-rich fractions that peaked at 20\%. None/0.4\% of the control samples have a gas-rich fraction lower than that of pRFE/cRFE galaxies. 
Similar trends are found for samples controlled in $M_{\rm BH}$ with the tolerance of $\Delta\log M_{\rm BH}/M_\odot<0.3$. 
We also constructed control RFI-Q samples with both NUV-r and stellar surface density controlled. The results are similar to those with NUV-r controlled and thus not presented here. 
Thus, the radio feedback that underpins the AI-based RFE/RFI classification is more important than $M_{\rm BH}$ and the NUV-r color in determining the amount of cold gas.

The analyzes so far have ignored the potential impact of misidentification. 
The intrinsic difference between RFE and RFI-Q galaxies should, therefore, be even larger. 
Assuming that the misidentification is insensitive to the HI content for quiescent galaxies, we evaluated the RFE fraction in gas-poor galaxies 
($\log M_{\rm HI}/M_\star<-1.5$). 
As shown in Fig.~\ref{fig:gf}B, the dependence on $M_\star$ becomes much weaker compared to the total population. 
The typical value of the RFE fraction at $\log M_\star/M_\odot>10.8$ is around 60\%, close to unity, although the uncertainty is large. 
This indicates that the majority of gas-poor galaxies are indeed RFE galaxies and most of RFI-Q galaxies are gas-rich. 

Since RFE galaxies on average reside in more massive halos than 
RFI galaxies of the same $M_\star$ (Fig.~\ref{fig:mps}), without radio heating RFE galaxies would contain a larger amount of HI gas because of the higher cooling rate in their more massive halos (\cite{Fabian2012}), which is in conflict with the observations presented here. 
This, of course, is the reason why we believe that heating sources are needed for RFE galaxies to counterbalance cooling. Furthermore, the predominance of RFE-off galaxies among the RFE population suggests that AGN feedback can effectively prevent gas cooling even after the AGN activity is paused. 
Therefore, despite its sporadic and anisotropic nature, radio feedback seems to be able to compensate the radiative loss that has much larger temporal and spatial scales than the feedback itself. These results give strong support to the hypothesis that feedback from strong radio AGNs can effectively heat the CGM, impeding gas cooling and maintaining the quiescence of host galaxies.  

However, radio heating is ineffective in the majority of massive quiescent (RFI-Q) galaxies. 
Many of these galaxies contain abundant cold gas (Fig.~\ref{fig:HI}), either cooled from the CGM or acquired from stripped satellite galaxies (\cite{Duc2011}), which is sustainable in the absence of strong radio heating. It is also plausible that the cold gas in some RFI-Q galaxies is the gas that existed prior to quenching. 
In the former case, the rotational axis of newly acquired gas is expected to be irregular or have some misalignment with that of the stellar component, whereas in the latter case, the left-over gas should show alignments with the stellar component. 

The ATLAS$^{3D}$ HI survey (\cite{Serra2012}) presented a HI survey of a volume-limited sample of 166 local early-type ATLAS$^{3D}$ galaxies. Each galaxy was observed for 12 hours using the Westerbork Synthesis Radio Telescope(WSRT). Of 166 galaxies, 53 galaxies are detected. 
The stellar masses of these galaxies are calculated based on their K-band luminosities using the method presented in \cite{Cappellari2013}.
We adopted the SFRs (\cite{Kokusho2017}) estimated by fitting the infrared SED from AKARI, WISE and 2MASS. The SFR for NGC 2594 is taken from another catalog (\cite{Leroy2019}). 
A large fraction of these galaxies are satellite galaxies located within massive groups and clusters, such as the Virgo cluster (\cite{Serra2012}). 
Therefore, the HI gas of the RFI-Q galaxies may be removed by environmental effects. We thus selected gas-rich quiescent galaxies, for which the environmental effect is expected to be weak, to represent RFI-Q galaxies. 
Only seven galaxies are quenched ($\log \rm{sSFR}<-11$) and possess a substantial amount of HI gas ($\log M_{\rm HI}/M_\star>-1.5$). These galaxies include NGC 6798, NGC 2594, NGC 2685, NGC 5173, UCG 09519, NGC 5582 and NGC 3073. Among them, NGC 6798, NGC 2685 and NGC 5582 were observed with WSRT for 108 hours per galaxy in follow-up observations. Recently, \cite{YangM2024} performed a detailed dynamical modeling on the stellar component and the HI gas for these galaxies and found that their HI velocity fields can be well fitted by rotational velocity fields.
We thus adopted the kinematic data provided by Serra et al. and Yang et al.

Serra et al. identified distinct signs of misalignment between the rotational axes of HI gas and stellar components in the first five galaxies. In the last one, complex morphology and unsettled kinematics for HI gas was noted. 
Yang et al. corroborated these findings, and, intriguingly, their detailed modeling also detected a considerable misalignment in NGC 5582. Thus, among the seven galaxies, five show evident misalignment signals, one presents an irregular morphology, and one remains uncertain, supporting that the HI gas in the RFI-Q galaxies is accreted post-quenching.
Except for the first, all of them lie in the SDSS region. We applied our AI method to their optical images and found that all of them are classified as RFI galaxies.
%Moreover, all seven galaxies have an HI size apparently larger than that of the stellar component (\cite{Serra2012}). It hints that the newly accreted gas has large angular momentum. 

% Why choose LERGs, Serra, what ai learns, galaxy fraction, power balance(maybe next )

%______________________________________________________________

\section{Summary and discussion}
\label{sec:conclusion}

In this paper, we have used multi-wavelength observations and AI classification method to investigate the role of radio-AGN feedback in maintaining the quiescence of massive galaxies. By distinguishing between galaxies with the potential for radio feedback effectiveness (RFE) and those where feedback is ineffective (RFI), we evaluate how often radio jets are responsible for preventing star formation. Our main conclusions are as follows:

\begin{enumerate}
    \item The application of artificial intelligence (AI) to classify galaxies based on optical images has proven to be an effective method for distinguishing between radio-feedback-effective (RFE) and radio-feedback-ineffective (RFI) galaxies, underscoring the potential of AI techniques in studying galaxy feedback mechanisms.
    \item The fraction of RFE galaxies increases with galaxy mass. Radio AGN feedback is effective in maintaining quiescence only in the most massive galaxies. However, only a small fraction of massive galaxies with $M_{\star}\sim10^{11}M_{\odot}$ remain quenched by this process.
    \item A comparison between RFE and RFI galaxies reveals that RFE galaxies typically reside in more massive halos and exhibit higher stellar mass, black hole mass, and concentration compared to RFI galaxies. Furthermore, RFE galaxies are dynamically hot, whereas quiescent RFI (RFI-Q) galaxies tend to possess extended cold-disk components.
    \item While RFE galaxies are generally deficient in cold gas, many quiescent RFI galaxies contain significant amounts of atomic gas. This suggests that radio feedback can effectively suppress gas cooling even after AGN activity ceases, whereas galaxies without strong AGNs may still harbor abundant cold gas despite being quenched.
   % \item The evolutionary paths of RFE and RFI galaxies are distinct: RFE galaxies maintain their quiescence through continuous suppression of gas cooling by AGN feedback, while RFI galaxies may resume star formation in the outer regions, potentially rejuvenating their stellar formations.
\end{enumerate}

Our study demonstrates that the feedback from strong radio AGNs can effectively maintain quiescence only in a small fraction of massive galaxies.
Because of significant differences in the radio activity, galaxy, halo and HI-gas properties, 
it is likely that RFI-Q galaxies have different evolutionary paths than RFEs, particularly 
in connection to the maintenance of quiescence.  For the gas-rich RFI-Q galaxies, radio feedback and any other heating mechanisms are expected to be ineffective.
One possible way for the gas-rich RFI-Q galaxies to maintain quiescence is that the cold gas has large angular momentum, so that the HI disk is too large to form stars and to feed the central black hole (\cite{Serra2012, LuS2022}). 
Indeed, all seven gas-rich and quenched galaxies in the ATLAS$^{\rm 3D}$ HI project have HI disks much larger than their stellar components (\cite{Serra2012}). 
Fig.~\ref{fig:HI} shows that many RFI-Q galaxies have HI masses similar to RFI-SF galaxies. 
This suggests that RFI-Q galaxies may resume their star formation (\cite{Zhou2021, Tanaka2024}), in contrast to RFEs which remain quenched. 
The small D4000 in the outskirts of some RFI-Q galaxies (Fig.~\ref{fig:mps}) may signify 
the onset of rejuvenation. These galaxies are currently forming stars in the outskirts dominated 
by newly accreted, high angular momentum gas in an extended distribution (\cite{Serra2012}).
It is thus likely that a fraction of star-forming galaxies are rejuvenated galaxies. 
This may explain the large ranges of concentration, outskirt dynamical states, and inner 
D4000 covered by RFI-SF galaxies in comparison to other types of galaxies (Fig.~\ref{fig:mps}). 
Some RFI-Q galaxies are deficient in cold gas. This suggests that
other mechanisms may heat the CGM and prevent the gas cooling 
for these galaxies. It is also possible that most of the CGM gas of these RFI-Q galaxies has already been consumed by star formation (\cite{ZhangZ2022}), as indicated by the large $M_\star/M_{\rm h}$ (Fig.~\ref{fig:mps}), so that the remaining CGM is too diffuse to cool. For example, in the elliptical galaxy NGC 720, only about $10^9M_\odot$ of the gas can cool over 5 billion years according to the prediction of a model without radio heating (\cite{Fabian2012}). 
%The diversity of galaxy evolution is much larger than the previous expectation. 

Many current galaxy formation models assume a radio (kinetic) mode of AGN feedback 
to heat the CGM and suppress gas cooling in massive galaxies 
(\cite{Croton2006, dave2019simba, Pillepich2018}). 
%or thermal feedback\cite{Schaye2015}
Such feedback is invoked to sustain the quiescence of almost all massive galaxies (\cite{LiH2025}). 
These models are thus at odds with our discovery that RFE galaxies constitute only a small fraction of all quenched massive galaxies (Fig.~\ref{fig:gf}) and that gas cooling may be effective in RFI-Q galaxies. 
One possible reason for the failure of these galaxy formation models is that the triggering of radio AGNs and feedback is not correctly modeled. 
Our results demonstrate that strong radio AGNs can be triggered mainly in entirely dynamically hot, massive galaxies that live in massive halos, in contrast to the assumption of many models that radio-AGN feedback becomes effective as soon as the mass of central supermassive black holes reaches some threshold (\cite{Weinberger2017, dave2019simba}). 
Clearly, some of the model assumptions need to be modified, and our results provide impetus for the search of a viable model.

Methodologically, we utilize noise-label learning to achieve scientific
discovery. To the best of our knowledge, this is the first time this method has
been applied in AI for Science. This approach enables robust classification for
science discovery from indirect or imperfect labels, which proves that this AI-based
knowledge discovery framework is suitable for any science problem that involves
label refinement or taxonomy. However, as shown in Fig. \ref{fig:AI_performance}, obtaining an accurate
classification is always a challenge when labels are incomplete. More work is certainly required to improve the performance of the classification. In the future, our method can be applied to higher-quality images of the upcoming Wide Field Survey Telescope survey (WFST, \cite{WangT2023SCPMA}) to improve its performance.

\begin{acknowledgements}
We thank Meng Yang and DanDan Xu for useful discussions. 
HYW is supported by the National Natural Science Foundation of China (NSFC, Nos. 12192224, 12595312), CAS Project for Young Scientists in Basic Research, Grant No. YSBR-062, the New Cornerstone Science Foundation through the XPLORER PRIZE, and the China Manned Space Project (No. CMS-CSST-2025-A04). 
YL and WLOY are supported by the JC STEM Lab of AI for Science and Engineering, funded by The Hong Kong Jockey Club Charities Trust, the Research Grants Council of Hong Kong (Project No. CUHK14213224). 
JW thanks support of research grants from  Ministry of Science and Technology of the People's Republic of China (No. 2022YFA1602902), NSFC (No. 12233001), and the China Manned Space Project (No. CMS-CSST-2025-A08).
ECW thanks support of the National Science Foundation of China (Nos. 12473008) and the Start-up Fund of the University of Science and Technology of China (No. KY2030000200).
HXZ acknowledges support from the National Key Research and Development Program of China (grant No. 2023YFA1608100), and from the NSFC (grant Nos. 12122303, 11973039).
YYC is funded by the China Postdoctoral Science Foundation (grant No. 2022TQ0329)
The authors gratefully acknowledge the support of Cyrus Chun Ying Tang Foundations. The work is supported by the Supercomputer Center of University of Science and Technology of China.
%The listed authors made substantial contributions to this manuscript; all co-authors read and commented on the document. HYW is responsible for the original idea, led the analysis and wrote the draft. HLL compiled and analyzed the observational data, developed the methods for purity and made the figures. YL developed and implemented the AI method and wrote the section for the AI method. HH compiled and analyzed the observational data.  HLL, YL and HH contribute equally to this work, and YL and HH are co-first authors. HJM contributed to the writing and interpretation of results. JW contributed to the analysis of HI data and interpretation of the results. WLOY contributed to the development of the AI method. ZWZ and QXL contributed to the analysis of weak lensing. ECW, HXZ, YYC, HL and MKZ contributed to the interpretation of the results.

\end{acknowledgements}

%------------------------------------------------------------------

\bibliographystyle{aa}
\bibliography{ref} 

\begin{appendix}
\section{What AI learns and its physical insight}

\begin{table*}
\caption{Numbers of galaxies in different classes based on the AI classification}   
\label{tab:AIresult}      
\centering          
\begin{tabular}{c c c c c} 
\hline\hline
 Initial label& \multicolumn{2}{c}{Q-LERGs} & \multicolumn{2}{c}{non-Q-LERGs}  \\ \hline
Full image& $\hat{y}_1>0.9$& $\hat{y}_1\leq0.9$& $\hat{y}_1>0.9$& $\hat{y}_1\leq0.9$\\
$\hat{y}_2>0.9$& 2668 & 421& 8786& 139\\ 
%10\% Test  Sample   & 365 & 9 & - & -  \\ \hline
$\hat{y}_2\leq0.9$& 4& 2& 30916& 357373\\\hline
P1($>0.75R_{50}$) & $\hat{y}_1>0.9$& $\hat{y}_1\leq0.9$& $\hat{y}_1>0.9$& $\hat{y}_1\leq0.9$\\
$\hat{y}_2>0.9$& 2668& 423& 7705(70.9\%)& 122\\ 
%10\% Test  Sample   & 365 & 9 & - & -  \\ \hline
$\hat{y}_2\leq0.9$& 3& 1& 32148(79.6\%)& 357239\\\hline
P2($>1.5R_{50}$) & $\hat{y}_1>0.9$& $\hat{y}_1\leq0.9$& $\hat{y}_1>0.9$& $\hat{y}_1\leq0.9$\\
$\hat{y}_2>0.9$& 2631& 463& 8058(69.4\%)& 115\\ 
%10\% Test  Sample   & 365 & 9 & - & -  \\ \hline
$\hat{y}_2\leq0.9$& 1& 0& 31090(76.8\%)& 357951\\\hline
C1($<0.75R_{50}$) & $\hat{y}_1>0.9$& $\hat{y}_1\leq0.9$& $\hat{y}_1>0.9$& $\hat{y}_1\leq0.9$\\
$\hat{y}_2>0.9$& 2560& 288& 32774(24.2\%)& 6245\\ 
%10\% Test  Sample   & 365 & 9 & - & -  \\ \hline
$\hat{y}_2\leq0.9$& 55& 192& 15715(32.4\%)& 342480\\\hline
C2($<1.5R_{50}$) & $\hat{y}_1>0.9$& $\hat{y}_1\leq0.9$& $\hat{y}_1>0.9$& $\hat{y}_1\leq0.9$\\
$\hat{y}_2>0.9$& 2661& 346& 21218(35.1\%)& 1746\\ 
%10\% Test  Sample   & 365 & 9 & - & -  \\ \hline
$\hat{y}_2\leq0.9$& 28& 60& 24999(52.8\%)& 349251\\\hline\hline
\end{tabular}
\tablefoot{The table shows the numbers of galaxies in different categories based on AI classification parameters $\hat{y}_1$ and $\hat{y}_2$. The ``Full image'' 
run is the fiducial run and uses the whole image of each galaxy for the training. The P1, P2, C1 and C2 runs use only part of each image(as indicated after the name) for the training.  Percentages shown in the fourth column indicate the fraction of galaxies in each $(\hat{y}_1, \hat{y}_2)$ category trained with partial images (P1, P2, C1, C2) that are also classified into the same $(\hat{y}_1, \hat{y}_2)$ category by the fiducial run. High percentage means that the corresponding run obtains result similar to that of the fiducial run.}
\end{table*}

To understand what information the AI technique acquires, we conducted four additional experiments in which the training utilized only portions of the image instead of the entire image for each galaxy. 
The initial two experiments focused on the peripheral regions beyond the radii of 0.75$R_{50}$ and 1.5$R_{50}$, respectively, which are referred to as P1 and P2.
The latter two experiments were limited to the central areas of each image, confined within the radii of 0.75$R_{50}$ and 1.5$R_{50}$, respectively.
The two experiments are referred to as C1 and C2, respectively. The results of these experiments are summarized in table~\ref{tab:AIresult}. 
Both the P1 and P2 experiments obtain results similar to those of the full image training, while there are many more non-Q-LERGs in C1 and C2 that have $\hat{y}_1\ge0.9$. 
These tests demonstrate that the galaxy outskirts are more crucial for the classification of RFE and RFI than the inner regions. It is consistent with the results shown in Fig.~\ref{fig:mps}: RFI often show extended cold-disk components in their outer regions, which contrasts with dynamically hot, centrally concentrated regions of RFE galaxies.

\section{Halo mass estimation using weak lensing}

\begin{figure*}
\includegraphics[width=\linewidth,keepaspectratio]{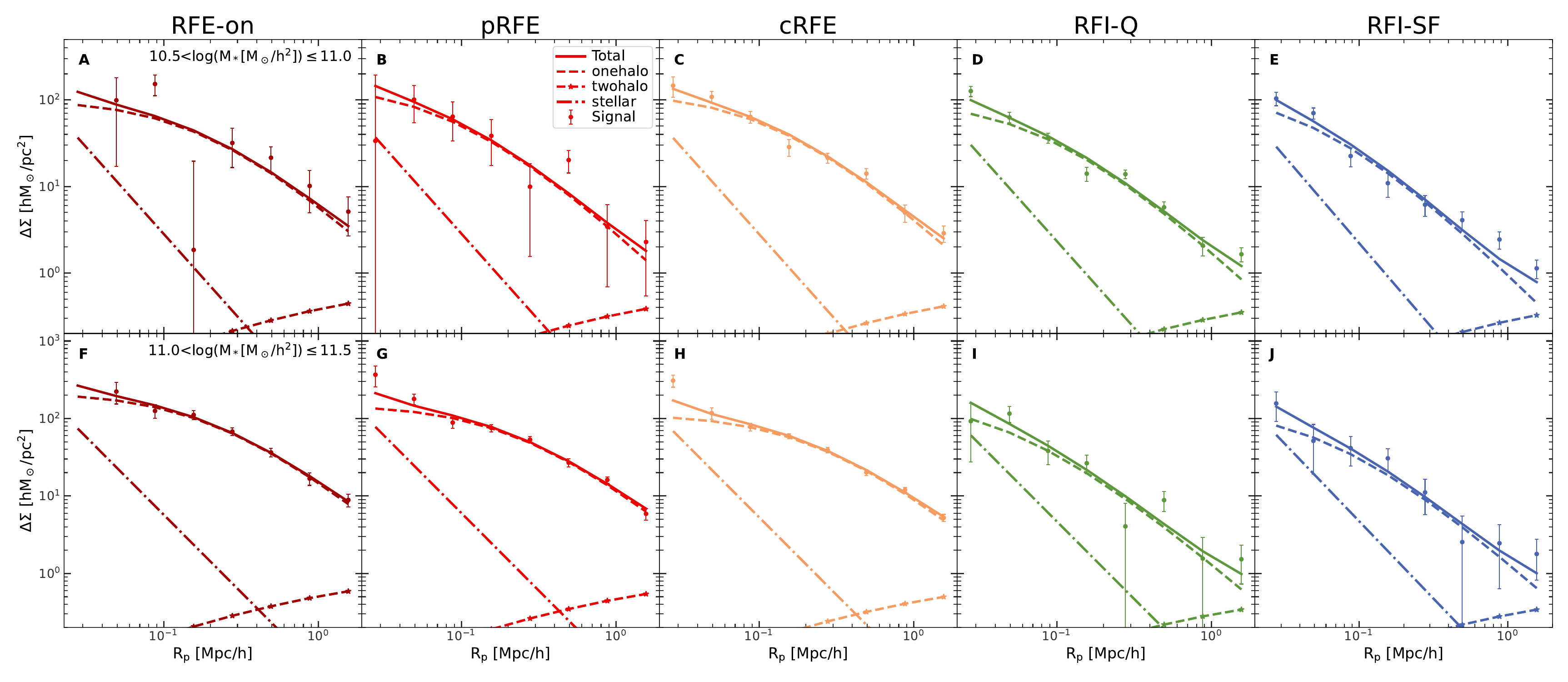}
\centering
\caption{{\bf Excess surface density from DECaLS and best fitting.} The symbols show the excess surface density (ESD) profiles obtained by stacking the lensing signal. Panel \textbf{A to E}: the results for galaxies with $10.5\le\log M_\star/M_\odot\le11.0$ in RFE-on, pRFE, cRFE, RFI-Q and RFI-SF samples, respectively. Panel \textbf{F to J}: the results for galaxies with $11.0\le\log M_\star/M_\odot\le11.5$. 
The shear catalog we used is constructed from the DECaLS DR8 imaging data. The error bars correspond to the standard deviation of 150 bootstrap samples.  We fit the ESD by using three components, stellar mass term (dash-dotted lines), one-halo term (dashed lines), and two-halo term (dotted lines with stars). The total fitting results are indicated by the solid lines.}\label{fig:lens}
\end{figure*}

In this section, we introduce the weak lensing technique to obtain the halo masses of our galaxy sample(see also \cite{ZhangZ2022}). The estimator, excess surface density (ESD, $\rm \Delta \Sigma$), at the projected radius R, defined as the mean surface density within a disk with radius R minus the boundary term
\begin{equation}
    \rm \Delta \Sigma (R)=\overline{\Sigma(<R)} - \Sigma(R)
\end{equation}
Additionally, it can be expressed in terms of the tangential shear
\begin{equation}
    \rm \Delta \Sigma (R)=\gamma_t(R)\Sigma_{\rm crit}
\end{equation}
Here,  $\gamma_t(R)$ represents the tangential component of the gravitational shear, which is related to the distortion of the source galaxy caused by the mass distribution of the lens galaxies. The critical surface density $\Sigma_{\rm crit}$ is a function of the lens and source redshifts, given by:
\begin{equation}
    \Sigma_{\rm crit} = \frac{c^2}{4\pi G}\frac{D_{\rm s}}{D_{\rm l} D_{\rm ls}}
\end{equation}
where $D_{\rm s}, D_{\rm l}, D_{\rm ls}$ are the angular distances to the source, to the lens, and between the lens and the source, respectively. We adopted the PDF-Symmetrization method developed by \cite{Zhang2017} to measure the ESD, as it effectively minimizes statistical uncertainties.  Fig.~\ref{fig:lens} shows the ESD for RFE-on, pRFE, cRFE, RFI-Q and RFI-SF galaxies in two overlap $M_\star$ bins. The error bars are estimated using 150 bootstrap samples.

Following previous studies (\cite{Mandelbaum2016, Luo2018ApJ, ZhangZ2022}), ESD is modeled using three components, 
\begin{equation}
    \rm \Delta \Sigma=\Delta\Sigma_{stellar}+\Delta\Sigma_{NFW}+\Delta\Sigma_{2h}
\end{equation} 
The first term is the contribution from the stellar mass of the central galaxies, which is modeled as a point mass. The second term is the contribution of the dark matter halo with a Navarro-Frenk-White density profile (\cite{Navarro1997}) which has two free parameters, mass $m_{\rm h}$ and concentration $c$.
Here,  $m_{\rm h}$ is the mass contained in the spherical region of
radius $r_{\rm 200m}$, where the mean mass density is equal to 200 times
the mean matter density of the Universe. 
Following \cite{Mandelbaum2016}, we do not consider the off-center effect, which is negligible according to our tests.
The third term is the two-halo term which was calculated by projecting along the line-of-sight of the halo-matter cross-correlation function $\xi_{\rm hm}$, where $\xi_{\rm hm}=\rm b(m_{\rm h})\xi_{\rm mm}$. Here $\rm \xi_{\rm mm}$ is the linear matter-matter correlation function and $\rm b(m_{\rm h})$ is the linear halo bias(\cite{Tinker2010}), these two are generated from \textit{COLOSSUS}. %In modeling the ESD, we only use the two halo term when the projected distance $r_{\rm p}$ from the lens galaxy is larger than the halo radius $r_{\rm 200m}$. 
We sample the posterior distributions of the two parameters $m_{\rm h}$ and $c$, using Markov Chain Monte Carlo (MCMC) with a software \textit{emcee} (\cite{Foreman-Mackey2013PASP}).
The best-fitting profiles are presented in Fig.~\ref{fig:lens}. 
The quoted mass, $m_{\rm h}$, is the median value of the posterior and its error bars indicate the 16 and 84 percentiles of the posterior.
We note that $m_{\rm h}$ actually accounts for the contribution of cold dark matter, diffuse gas, and satellites around the centrals but does not include the contribution of central galaxies, which are modeled separately. So, the total mass of the halo should include all components within the virial radius of the halo, is $M_{\rm h}=m_{\rm h}+M_\star$. Usually, $M_\star\ll m_{\rm h}$, so $M_{\rm h}$ is very close to $m_{\rm h}$. In this paper, when we talk about halo mass, what we mean specifically is the total mass of the halo, i.e. $M_{\rm h}$.

\end{appendix}

\end{document}